\documentclass{LMCS}

\def\dOi{9(4:14)2013}
\lmcsheading%
{\dOi}
{1--21}
{}
{}
{Jun.~\phantom01, 2011}
{Nov.~14, 2013}
{}

 \ACMCCS{[{\bf Mathematics of computing}]:
  Continuous mathematics---Topology---Point-set topology}

\usepackage{amssymb}
\usepackage{amsmath}
\usepackage{latexsym}
\usepackage{hyperref}

\allowdisplaybreaks

 \newcommand{\hideb}[1]{}

\def \cal{\mathcal}

\def \dom{{\rm dom}}

\def \In{{\subseteq}}
\def \om{{\Sigma^\omega }}
\def \pf{:\hspace{0.6ex}\subseteq \hspace{-0.4ex}}

\def \range{{\rm range}}
\def \s{{\Sigma^*}}
\def \xx{{\bf X}}
\def \yy{{\bf Y}}
\def \zz{{\bf Z}}

\def \tp {{\hspace{-.4ex}\widetilde{\hspace{.3ex}\psi}\hspace{.1ex}}}
\def \tk{ {\widetilde\kappa}}

\def\IN{{\mathbb{N}}}
\def\IN{{\mathbb{N}}}

\def\IR{{\mathbb{R}}}
\def\IR{{\mathbb{R}}}

\newcommand{\an}{\ \wedge\ }
\newcommand{\mto}{\rightrightarrows}

\newcommand{\nufs}{{       \nu^{\rm fs}}}

\newcommand{\bb}{ \hspace{-0.76ex}- \hspace{-0.80ex}}

\newcommand{\ETS}{\mbox{$\mathcal T$}\hspace{-.5ex}}
\newcommand{\ETSP}{{\mbox{$\mathcal{ T\hspace{-.4ex}P}$}}}
\newcommand{\ETSS}{{\mbox{$\mathcal{ T\hspace{-.4ex}S}$}}}
\newcommand{\ETSO}{{\mbox{$\mathcal{ T\hspace{-.4ex}O}$}}}
\newcommand{\ETSC}{{\mbox{$\mathcal{ T\hspace{-.4ex}C}$}}}

\begin{document}

\title[Computable Tychonoff]{Products of effective topological spaces and a uniformly computable Tychonoff Theorem}

\author[R. Rettinger]{Robert Rettinger}   
\address{Dpt. of Mathematics and Computer Science, University of Hagen, Germany} 
\email{\{Robert.Rettinger, Klaus Weihrauch\}@FernUni-Hagen.de}  

\author[]{Klaus Weihrauch} 
\address{\vspace{-18 pt}}    


\begin{abstract}

This article is a fundamental study in computable analysis. In the framework of Type-2 effectivity, TTE,  we investigate computability aspects on finite and infinite products of effective topological spaces. For obtaining uniform results we introduce natural multi-representations of the class of all effective topological spaces, of their points, of their subsets and of their compact subsets. We show that the binary, finite and countable product operations on effective topological spaces are computable. For spaces with non-empty base sets the factors can be retrieved from the products.
We study computability of the product operations  on points, on arbitrary subsets and on compact subsets. For the case of compact sets the results are uniformly computable versions of Tychonoff's Theorem (stating that every Cartesian product of compact spaces is compact) for both, the cover multi-representation and the ``minimal cover'' multi-representation.
\end{abstract}


\maketitle

\section{Introduction}\label{seca}

In this article we study basic aspects of computable analysis in the framework of  Type-2 theory of effectivity (TTE) \cite{Wei00,BHW08,WG09}.
In computable analysis usually computability has been studied on fixed computable structures such as computable topological spaces (e.g. $\IR^n$), computable metric spaces, computable Banach spaces, computable Hilbert spaces, computable Sobolev spaces or computable measure spaces.
Computability of such a structure means that some of its ``characteristic data'' can be computed.

Sometimes in a proof,  an ``intermediate'' structure, for example a metric space, is used the characteristic data of which can be computed from not necessarily computable input data and hence may be non-computable. Therefore, the known theorems about computable metric spaces cannot be applied. A more general computability theory uniform on all metric spaces is needed where the metric space occurs as a parameter and the functions in the theorems are computable also in the characteristic data of the metric space.

Often the validity of such uniform computability results is almost obvious and used
in a somewhat informal fashion. In some articles proofs of the uniform versions are presented. But sometimes the validity of the uniform version is not at all obvious.
For example, to prove the computability of bi-holomorphic mappings on simply connected domains, a computable Tychonoff theorem is used to prove in a simple way the compactness of certain function spaces
\cite{Ret11}. However, without a uniform version of this theorem, the results either depend on some kind of informal argumentation on uniformity, or
are restricted to a very bounded class of domains. In this article we will prove, among others, a uniformly computable Tychonoff theorem.

Since the cardinality of the class of underlying spaces is usually greater than that of the continuum (that is, the set of infinite sequences of symbols), it has no representation.
Sometimes the cardinality problem can be solved by factorization, where spaces are identified which have the same data specifying computability. In this way one gets a class of at most continuum cardinality (see e.g. \cite{GM09},\cite{Re07}\cite[Section~8.1]{Wei00}).
In our case this method fails. We solve the problem by using a multi-representation of the class of all ``effective'' spaces under consideration.

In this article we continue the study of elementary computable topology \cite{Wei00,WG09,Wei10,Wei13,GSW07,Sch03}. We define a natural multi-representation $\Delta$ of the class of all effective topological spaces \cite{WG09} and study the product operation on this class.
We work in the representation model of computable analysis \cite{KW85,Wei00, BHW08}.

In Section~\ref{secb} we introduce some basic definitions and notations from computable analysis. For more details see \cite{Wei00,BHW08,WG09}.

In Section~\ref{secc} we define a multi-representation $\Delta$ of the class of all  effective topological spaces, where we apply the definition of ``effective topological space'' from  \cite{WG09}.
We mention that there are other slightly different  definitions of ``effective topological spaces'', e.g. in \cite{Wei00}, which, however, have turned out to be less natural and useful.
We formulate a meta-theorem by which essentially all theorems in \cite{WG09} stating computability have a computable version uniform in the spaces under consideration.
The  canonical (multi-)representations $\delta$ of the points, $\theta$ of the open sets, $\tp$ of all subsets and $\kappa$ and $\tk$ of the compact sets for a fixed effective topological space from \cite{WG09} are generalized in two ways to the class of all spaces.

In Section~\ref{secd} we define finite and infinite products of effective topological spaces. We characterize the product by universal properties.
We prove that the product operations on the spaces are computable w.r.t. the multi\bb representation~$\Delta$. For spaces with non-empty base sets the factors can be retrieved from their products. In general the product is, up to equivalence of spaces, commutative and associative.

In Section~\ref{sece} we study computability of the product operations and their inverses on points (a tuple of points from a sequence of spaces is mapped to a point in the product space), on arbitrary sets and on compact sets for finite and infinite sequences of effective topological spaces. We prove computability uniform in the class of all effective topological spaces. As corollaries we obtain the versions for fixed computable spaces and for computable points of fixed computable spaces.

By Tychonoff's theorem from topology, every Cartesian product of compact spaces is compact. As a main result we obtain that the (finite as well as countable) product of compact subsets of effective topological spaces can be computed uniformly in the spaces. This is true for the multi-representations of the compact sets by finite covers as well as for the multi-representations of the compact sets by minimal finite covers \cite[Section~5.2]{Wei00},\cite{WG09}.

Brattka \cite{Bra08b} has shown that $\prod_{i=1}^{\infty}[-|x_i|;|x_i|]$ is a computable compact set in $\IR^\IN$ if $(x_i)_{i\in\IN}$ is a computable sequence of real numbers.
Gherardi et al. \cite[Lemma~8.8]{GM09} have shown that the operator
$(x_i)_{i\in\IN}\mapsto \prod_{i=1}^{\infty}[-|x_i|;|x_i|]$ is computable. These results are applications of  Theorem~\ref{t6}. Escard\'o \cite{Esc08} has proved that
the computable countable product of searchable subsets of a domain $D$ is searchable in $D^\IN$, where searchable sets are a special kind of computable compact sets. This corresponds to the following corollary of Corollary~\ref{c5}(\ref{c5c}): For a computable topological space the product of  a computable sequence of compact sets is a computable compact set (in the computable product space). Since ``searchable'' and  ``computably compact'' are different concepts, the two results are incomparable. The main result in this article, Theorem~\ref{t6}(\ref{t6c}), is much more uniform with respect to computability.

In this article we study computability on the class of all effective topological spaces.
The methods  can be used  as a blueprint to provide uniform computability on many other classes of spaces considered in Type-2 theory of effectivity.

\section{Preliminaries}\label{secb}
In this section we recall some definitions of Type-2 theory of effectivity (TTE). We nevertheless assume basic knowledge on the theory and furthermore depend
on the notations introduced in \cite{Wei00,BHW08,WG09}.

Let $\Sigma$ be a finite alphabet such that $0,1\in\Sigma$. By $\s$ we denote the set of
finite words over $\Sigma$ and by $\om$ the set of infinite sequences
$p:\IN\to\Sigma$ over $\Sigma$, $p=(p(0)p(1)\ldots)$.
We use the  ``wrapping function''
$\iota:\s\to\s$, $\iota(a_1a_2\ldots a_k):=110a_10a_20\ldots a_k011$
for coding words such that $\iota(u)$ and $\iota(v)$ cannot overlap properly.
Let $\langle i,j\rangle:=(i+j)(i+j+1)/2+j$ be the bijective Cantor pairing function on $\IN$.
We consider standard functions for finite or countable tupling on $\s$ and $\om$
\cite[Definition~2.1.7]{Wei00}, in particular,
\begin{eqnarray*}
\langle u_1,\ldots, u_n\rangle &:=&\iota(u_1)\ldots\iota(u_n)\,,\\
\langle u,p\rangle & :=&\iota(u)p\,,\\
\langle p_1,p_2,\ldots,p_n\rangle&:=& (p_1(0)p_2(0)\ldots p_n(0)p_1(1)p_2(1)\ldots p_n(1))p_1(2)\ldots\,,\\
\langle p_0,p_1,\ldots\rangle \langle i,j\rangle&:=&p_i(j)
\end{eqnarray*}
where $u,u_1,u_2,\ldots\in\s$ and  $p,q,p_0,p_1,\ldots\in\om$.
For $u\in\s$ and $p\in\s$ or $p\in\om$, $u\ll p$ means that $\iota(u)$ is a subword of $p$ (that is, $p=v\iota(u)q$ for some $v,q$). As a technical detail, notice that  $n$ can be determined from $\langle u_1,\ldots, u_n\rangle$ in Line 1 but not from $\langle p_1,p_2,\ldots,p_n\rangle$ in Line 3.

For a notation (that is, a surjective function) $\mu\pf\s\to Y$ the canonical notation $\mu^{\rm fs}$ of the finite subsets of $Y$ is defined by
 $\mu^{\rm fs}(w) =W$  iff
$(\forall v\ll w)\,v\in \dom(\mu)$ and  $W=\{\mu(v)\mid v\ll w\}$
\cite{Wei00,WG09}.
For the natural numbers we will use the notation $\nu_\IN$, where $\nu_\IN(0^n):=n$ and $w\not\in\dom(\nu_\IN)$ for all other words $w\in\s$. Then $\nu_\IN$ is equivalent to other standard notations of $\IN$ \cite{Wei00}.

In TTE representations are used as ``naming systems'' for sets of abstract objects and computations are performed on ``names'' from $\s$ or $\om$. In this article multi-representations are essential. Formally, a multi-representation of a class (not necessarily set) $M$ is a relation $\delta\subseteq Y\times M$ where $Y=\s$ or $Y=\om$ such that $(\forall x\in M) (\exists p\in Y)(p,x)\in\delta$. We write $\delta:Y\mto M$ and define $\delta(p):=\{x\in M\mid (p,x)\in\delta\}$ and $\dom(\delta):= \{ p\in Y\mid\delta(p)\neq\emptyset\}$.
We do not consider $\delta$ as a (single-valued) representation of a subset of $2^M$. If $x\in\delta(p)$ we can say ``$p$ is a $\delta$-name of $x$''. In general such a name does not identify an object but only gives some property of the object. (For example, ``Peter'' is the first name of many people.)
We mention that  in TTE there are two interpretations of the concept ``multi-function'' which can be distinguished formally by the definition of composition, see \cite[Sections 3 and 6]{Wei08}.

Computability on multi-represented sets is defined as follows.
Let $\gamma: Y\mto M$ and $\gamma': Y'\mto M'$ ($Y,Y'\in\{ \Sigma^\ast,\Sigma^\omega\}$) be multi-representaions of classes
$M$ and $M'$, \, respectively. A partial function $g\pf Y\rightarrow Y'$ realizes $f\pf M\to M'$ if $f(x)\in \gamma'\circ g(p)$ whenever $x\in\dom(f)\cap\delta(p)$. The partial function $f\pf M\to M'$ is $(\gamma,\gamma')$-computable iff there exists a computable (by a Type-2-Turing machine) function $g\pf Y\rightarrow Y'$ which realizes $f$. A subset $X\In M$ is $\gamma$-r.e. (recursively enumerable), iff there is a Type-2 machine $N$ such that for all $x,p$ with $x\in\gamma(p)$:
$N$ halts on input $p$ iff $x\in X$ \cite{WWD09}.

Computability on products can be defined in the same way. See e.g. \cite[Section~6]{Wei08}\ for further details.
Given two multi-representations $\gamma$, $\gamma'$ of classes $M\subseteq M'$, respectively, we say that $\gamma$ is reducible to $\gamma'$ ($\gamma\leq\gamma'$ for short) iff the inclusion $m\mapsto m$ of $M$ into $M'$ is $(\gamma,\gamma')$-computable. We call $\gamma$ and $\gamma'$ equivalent ($\gamma\equiv \gamma'$) iff $M=M'$ and
$\gamma\leq\gamma'$ and $\gamma'\leq\gamma$. Notice that $\gamma\leq\gamma'$ iff $\gamma(p)\subseteq \gamma'\circ h(p)$ for some computable function $h$. (If $\gamma$ and $\gamma'$ were considered as single-valued representations of $2^M$ and $2^{M'}$, respectively, then we should use equality $\gamma(p)=\gamma'\circ h(p)$.)

Let $\nu:\s\mto X$ be a multi-representation. Define $\delta_\nu : \om\mto X$ by $\delta(\iota(w)00\ldots)=\nu(w)$ for $w\in\dom(\nu)$.
 Since the function $h:\s\to\om$, $h(w):=\iota(w)00\ldots$ and its inverse are computable
\cite[Theorem~2.1.8]{Wei00}, $\nu\equiv\delta_\nu$ where the functions $h^{-1}$ and $h$ translate back and forth. Notice that the same function $h$ works for all notations. Therefore, for convenience it suffices to consider only multi-representations $\delta:\om\mto X$ in all theorems where multi-representations can be replaced by equivalent ones.

The functions that are computable w.r.t. multi-representations are closed under composition \cite[Sections 3 and 6]{Wei08}. More generally, they are closed under programming with ``Turing machines on represented sets'' \cite{TW11b}, which are a useful model for discussing algorithms in computable analysis. Implicitly we will use this model without further mentioning.

\section{Computability on the Class of Effective Topological Spaces}\label{secc}
The basic structure in \cite{WG09} is the computable topological space.

\begin{defi}[effective/computable topological space \cite{WG09}]\label{d1}
An {\em effective topological 
space} is a 4-tuple $\xx=(X,\tau,\beta,\nu)$
such that $(X,\tau)$ is a topological $T_0$-space and
$\nu\pf\s\to \beta$ is a notation of a base $\beta$ of $\tau$.
Let $\mathcal T$ be the class of all effective topological spaces.

$\xx$ is a {\em computable} topological space if $\dom(\nu)$ is recursive and
\begin{eqnarray}\label{f30}
\nu(u)\cap\nu(v)=\bigcup\{\nu(w)\mid (u,v,w)\in S\} \ \
\mbox{for all}\ \ u,v\in\dom(\nu)
\end{eqnarray}
for some r.e. set $S\In(\dom(\nu))^3$.
\end{defi}

A closer look at \cite{WG09} shows that all the proofs of computability use from the underlying computable topological space
only the characteristic function of $\dom(\nu)$ and an enumeration of a set $S\In(\dom(\nu))^3$ such that (\ref{f30}) holds.
(Spaces with the same characteristic function and the same enumeration cannot even be distinguished.)
Therefore, the whole theory can be generalized to effective topological spaces where the formerly computable functions become computable with the (not necessarily computable) characteristic function of $\dom(\nu)$ and some (not necessarily computable) enumeration of the set $S$ as oracles.
Following these ideas we introduce a multi-representation of the class of effective topological spaces as follows.

\begin{defi}\label{dt1} Define a multi-representation $\Delta:\om\mto { \ETS}$ of the class $\bf \ETS$ of effective topological spaces as follows:
$\xx=(X,\tau,\beta,\nu)\in \Delta\langle r,s\rangle$ ($r,s\in\om$) iff $r$ enumerates the graph of the characteristic function of $\dom(\nu)$ and $s$ enumerates a subset $S\In(\dom(\nu))^3$ such that
\begin{eqnarray}\nonumber
\nu(u)\cap\nu(v)=\bigcup\{\nu(w)\mid (u,v,w)\in S\} \ \
\mbox{for all}\ \ u,v\in\dom(\nu)\,.
\end{eqnarray}
\end{defi}

Obviously, $\bf X$ is a computable topological space, iff $\xx\in\Delta(t)$ for some computable $t\in\om$.
For every set $S\In (\s)^3$ let ${\cal T}_S$ be the class of effective topological spaces  for which $S$ realizes intersection.
Then  every non-empty class ${\cal T}_S$
has a maximal element ${\bf X'}\in{\cal T}_S$ such that $\xx\in {\cal T}_S$ iff
  $\xx$ can be obtained from $\bf X'$ by deleting some points and renaming the other points  \cite[Proposition~34, Theorem~36]{WG09}. In particular, the representation $\Delta $ is multi-valued in general, and spaces with the same name may be not even homeomorphic. Our observation on uniformization can be formulated as follows.

\begin{thm}[Meta-Theorem]\label{t1}
Consider a theorem in \cite{WG09} of the following form where $Y_0,\ldots,Y_n\in\{\s,\om\}$ :
\begin{quote}
Let $\xx=(X,\tau,\beta,\nu)$ be a computable topological space. Then there is a computable function $h\pf Y_1\times\ldots\times Y_n\to Y_0$ such that ${\cal Q}(h,\xx)$.
\end{quote}
Then the following uniform generalization is true:
\begin{quote}
There is a computable function $\overline h \pf \om \times Y_1\times\ldots\times Y_n\to Y_0$ such that ${\cal Q}( h_t,\xx)$ for all \, $\xx=(X,\tau,\beta,\nu)\in\Delta(t)$, where $h_t(y_1,\ldots,y_n):=\overline h(t,y_1,\ldots,y_n)$.
\end{quote}
This meta-theorem holds accordingly if finitely many effective topological spaces are involved.
\end{thm}

We can say: There is a function $h$ uniformly computable in $\bf X$ such that ${\cal Q}(h,\xx)$.\\

\proof  For every theorem  in in \cite{WG09} under consideration check its proof and observe that from every computable topological space $\xx=(X,\tau,\beta,\nu)$
only the characteristic function of $\dom(\nu)$ and an enumeration of a set $S\In(\dom(\nu))^3$ such that (\ref{f30}) holds are used.
\qed

For effective topological spaces the following natural (multi)-representations have been introduced \cite{KW87,Wei00,WG09,Sch03}.

\begin{defi}\label{d4}
For an effective topological space $\xx=(X,\tau,\beta,\nu)$ define a representation $\delta_\xx$ of the points, a representation $\theta_\xx$ of the open sets, a multi-representation $\tilde \psi_\xx$ of all subsets and   multi-representations $\kappa_\xx$ and $\tk_\xx$ of the compact subsets as follows:
\begin{eqnarray}
\label{f4} x=\delta_\xx(p) &:\iff &(\forall w\in \s)\,( w\ll p\iff x\in\nu(w))\,,\\
\label{f5} W=\theta_\xx(p) &:\iff&
\left\{
\begin{array}{l}
 w\ll p\Longrightarrow w\in\dom(\nu),\\
W=\bigcup\{\nu(w)\mid   w\ll p\},
\end{array}\right.\\
 \label{f6} B\in\tp_\xx(p) &:\iff &
(\forall w\in\s)\,( w\ll p\iff B\cap\nu(w)\neq \emptyset)\,,\\
\label{f7} K\in\kappa_\xx(p)&:\iff&
(\forall w\in\s)\,( w\ll p\iff K\In\bigcup\nufs(w))\,,\\
\label{f7a}K\in \tk_\xx(p)&\iff& (\forall w\in\s)\,( w\ll p \iff \left\{
\begin{array}{c}  K\In\bigcup\nufs(w)) \an \\
(\forall u\ll w)\nu(u)\cap K\neq\emptyset\,.
\end{array}\right.
\end{eqnarray}
\end{defi}

In (\ref{f4}) ``$x\in\nu(w)$'' includes $w\in\dom(\nu)$, correspondingly in
(\ref{f6}) and $w\in \dom(\nufs)$ in (\ref{f7}) and (\ref{f7a}).
The above definitions induce mappings from \ETS\ to the class of multi-representations of points, open sets, subsets and compact sets ($\xx\mapsto \delta_\xx$ etc.). Every such mapping can be generalized to a multi-representation as follows.
\begin{defi}\label{dt2}
Let $\ETSP$, $\ETSO$, $\ETSS$ and $\ETSC$ ({\bf p}oints, {\bf o}pen sets, arbitrary {\bf s}ets and {\bf c}ompact sets) be the class of pairs $(\xx,x)$, $(\xx,W)$, $(\xx,B)$ and $(\xx,K)$, respectively, where $\xx\in\ETS$, $x\in X$ is a point, $W\subseteq X$ is an open set, $B\subseteq X$ is a set, and $K\subseteq X$ is a compact set.
Define multi-representations $\delta$, $\theta$, $\tp$, $\kappa$ and $\tk$ of $\ETSP$, $\ETSO$, $\ETSS$ and $\ETSC$, respectively, as follows:
\begin{eqnarray*}
(\xx,x)\in \delta(s) &:\iff & x=\delta_\xx(s)\,,\\
(\xx,W)\in \theta(s) &:\iff & W=\theta_\xx(s)\,,\\
(\xx,B)\in {\tp}(s) &:\iff & B\in\tp_\xx(s)\,,\\
(\xx,K)\in \kappa(s)&:\iff & K\in\kappa_\xx(s)\,,\\
(\xx,K)\in \tk(s)&:\iff & K\in\tk_\xx(s)\,.
\end{eqnarray*}
for $\xx\in\ETS$, $x\in X$, open $W\In X$, $B\In X$, and compact $K\In X$.
\end{defi}

Computability with respect to the above multi-representations means that the realizing function is independent of the represented effective $T_0$-space. To allow to use information on this space we introduce a second kind of multi-representations, again derived from the natural multi-representations $\delta_\xx$, $\theta_\xx$, $\tp_\xx$ and $\kappa_\xx$.

\begin{defi}\label{dt2b}
Define multi-representations $\delta^\Delta$, $\theta^\Delta$, ${\tp}^\Delta$,
$\kappa^\Delta$ and $\tk^\Delta$  of $\ETSP$, $\ETSO$, $\ETSS$ and $\ETSC$, respectively, as follows:
\begin{eqnarray*}
(\xx,x)\in \delta^\Delta\langle r,s\rangle&:\iff &\xx\in\Delta(r)\an x=\delta_\xx(s)\,,\\
(\xx,W)\in \theta^\Delta\langle r,s\rangle&:\iff &\xx\in\Delta(r)\an W=\theta_\xx(s)\,,\\
(\xx,B)\in {\tp}^\Delta\langle r,s\rangle&:\iff &\xx\in\Delta(r)\an B\in\tp_\xx(s)\,,\\
(\xx,K)\in \kappa^\Delta\langle r,s\rangle&:\iff &\xx\in\Delta(r)\an K\in\kappa_\xx(s)\,,\\
(\xx,K)\in \tk^\Delta\langle r,s\rangle&:\iff &\xx\in\Delta(r)\an K\in\tk_\xx(s)\,.\end{eqnarray*}
for $\xx\in\ETS$, $x\in X$, open $W\In X$, $B\In X$, and compact $K\In X$.
\end{defi}

Other multi-representations defined in \cite{WG09} can be generalized accordingly.
Notice that
\begin{eqnarray}\label{f9}(\xx,x)\in\delta(p)\iff x=\delta_\xx(p)
\iff \{x\}\in\tp_\xx(p)\iff (\xx,\{x\})\in\tp(p)\,,
\end{eqnarray}
hence $\delta$ can be considered as the restriction of $\tp$ to the sets with cardinality~$1$ (correspondingly for $\delta^\Delta $ and $\tp^\Delta$).
If $K\in\tk_\xx(p)$, then $p$ is a list of all finite sets $\{U_1, \ldots, U_n\}$ of base sets such that $K$ is contained in their union and every $U_i\in\beta$ intersects $K$.
This allows us to derive $\tk$ from $\kappa$ and $\tp$. For multi-representations $\gamma:\om\mto X$ and $\delta:\om\mto Y$ the conjunction
$\gamma\wedge\delta:\om\mto X\cap Y$
is defined as follows (\cite{Wei00,Sch03,WG09}:
\begin{eqnarray}\label{f10}
(\gamma\wedge\delta)\langle p,q\rangle&:=&\gamma(p)\cap\delta(q)\,.
\end{eqnarray}

The two multi-representations of the compact sets are related by $\tp$ as follows.

\begin{lem}\label{l1}\hfill
\begin{enumerate}
\item\label{l1a} For every effective topological space $\xx$, \ $ \kappa_\xx\wedge \tp_\xx\leq \tk_\xx$,\\
 for every computable topological space $\xx$, \ $ \kappa_\xx\wedge \tp_\xx\equiv \tk_\xx$,

\item\label{l1b} $\kappa\wedge \tp\leq \tk\,,$
\item\label{l1c}
$\tk^\Delta\equiv \kappa^\Delta\wedge \tp^\Delta\,.$
\end{enumerate}
\end{lem}

\proof

(\ref{l1b}) suppose $(\xx, K)\in(\kappa\wedge \tp)\langle p_1,p_2\rangle$.
Then $K\in\kappa_\xx(p_1)$ and $K\in\tp_\xx(p_2)$.
From the list $p_1$ of all finite base-covers of $K$ and the list $p_2$ of all base elements $U$ with $U\cap K\neq\emptyset$ we can compute a list $s$ of all minimal finite base-covers of $K$, hence $K\in\tk_\xx(s)$. Therefore, $(\xx,K)\in\tk( s)$, and
$\kappa\wedge \tp\leq \tk$.

(\ref{l1c}) Suppose $(\xx,K)\in\tk^\Delta\langle r,s\rangle$. Then $\xx\in\Delta(r)$ and $K\in\tk_\xx(s)$. From $\dom(\nu_\xx)$, hence form $r$,  and the list $s$ of all minimal finite  base-covers we can compute a list of all finite base-covers, hence a function $p_1$ such that $K\in\kappa_\xx(p_1)$, hence $(\xx,K)\in \kappa^\Delta\langle r,p_1\rangle$. From $s$ we can compute a list of all base elements $U$ such that $K\cap U\neq\emptyset$, hence a function $p_2$ such that $K\in\tp_\xx(p_2)$, hence $(\xx,K)\in\tp^\Delta\langle r,p_2\rangle$.
Then $(\xx, K)\in(\kappa^\Delta\wedge \tp^\Delta)\langle \langle r,p_1\rangle,\langle r,p_2\rangle\rangle$. Therefore, $\tk^\Delta\leq \kappa^\Delta\wedge \tp^\Delta$.

$(\kappa^\Delta\wedge \tp^\Delta)\leq \tk^\Delta$ follows straightforwardly from (\ref{l1b}).

(\ref{l1a}) The first statement follows from  (\ref{l1b}) and the second statement from (\ref{l1c}).
\qed

For translating $\tk_\xx$ to $\kappa_\xx$ from a list of all minimal finite base-covers we must find a list of all finite base-covers. But this cannot be done without knowing $\dom(\nu_\xx)$. Therefore, $\tk_\xx \leq \kappa_\xx\wedge \tp_\xx$ is false in general and $\tk\leq \kappa\wedge \tp$ is false. If we replace the class $\ETS$ by the subclass $\mathcal T'$ of spaces $\xx$ such that $\nu_\xx$ is a total function, then $\tk\leq\kappa\wedge \tp$. We do not know whether the restriction to $\mathcal T'$ is sufficiently general in future applications.

\section{Products of Spaces}\label{secd}

We generalize the definitions of the representations $[\delta_1,\ldots,\delta_n]$, $[\delta]^n$ and $[\delta]^\IN$ of products and sums for single-valued representations introduced in
\cite[Definitions~3.3.3,~3.3.14]{Wei00}  to multi-representations. Remember that the disjoint union of a sequence $(A_n)_{n\in\IN}$ of sets is defined by $\biguplus_n A_n:=\{(n,x)\mid x\in A_n\}=\bigcup_n \{n\}\times A_n$.

\begin{defi}\label{d3} Let $\delta_i:\om\mto X_i$ ($i=1,2,\ldots$) be multi-representations.

\begin{enumerate}
\item \label{d3a} Define a multi-representation of the finite product $X_1\times\ldots\times X_n$ by
\[[\delta_1,\ldots,\delta_n]\langle p_1,\ldots,p_n\rangle=\delta_1(p_1)\times\ldots\times \delta_n(p_n)\]

\item \label{d3b} Define a multi-representation of $\,\biguplus_{n\geq 1}X_n$ by
\[(\delta_1\vee \delta_2\vee\ldots\,)(1^n0 p):=\{n\}\times \delta_n(p)\,.\]

\item \label{d3c} Define a multi-representation $[\delta_1,\delta_2,\ldots\,]^+$ of the disjoint union of finite products \\ $\biguplus_{n\geq 1}  X_1\times\ldots\times X_n$ by
\[[\delta_1,\delta_2,\ldots\,]^+(1^n0\langle p_1,\ldots,p_n\rangle):=\{n\}\times
\delta_1(p_1)\times\ldots\times \delta_n(p_n)\,.\]

\item \label{d3d} Define a multi-representation of the infinite product $X_1\times X_2\times\ldots$ by

$(x_1,x_2,\ldots\,)\in[\delta_1,\delta_2,\ldots\,]\langle p_1,p_2,\ldots\,\rangle \iff (\forall\;i\geq 1)\, x_i\in\delta_i(p_i)$.
\end{enumerate}
Let $[\delta]^n:=[\delta,\ldots,\delta]$ ($n$-times),
$[\delta]^+:=[\delta,\delta,...\,]^+$ and $[\delta]^\IN:=[\delta,\delta,...\,]$.
\end{defi}

The following three characterizations show that the product $[\delta_1,\delta_2]$ is very natural.

\begin{thm}\label{t4}
 Let $\delta_i:\om\mto X_i$ ($i=1,2$) be multi-representations. For  multi-representations $\gamma:\om\mto X_1\times X_2$ of $X_1 \times X_2$ define
 \begin{eqnarray*}
 S(\gamma) & :\iff & \mbox{the function $(x_1,x_2)\mapsto (x_1,x_2)$ is $(\delta_1,\delta_2,\gamma)$-computable\,,}\\
 A(\gamma) &:\iff & \left\{\begin{array} {lll}
 \mbox{ $(x_1,x_2)\mapsto x_1$ is $(\gamma,\delta_1)$-computable}\ \mbox{ and}\\
 \mbox{ $(x_1,x_2)\mapsto x_2$ is $(\gamma,\delta_2)$-computable}\,.
 \end{array}\right.
 \end{eqnarray*}
Then
\begin{eqnarray}
\label{f11} [\delta_1,\delta_2]\leq \gamma & \iff & S(\gamma)\,,\\
\label{f12} \gamma\leq [\delta_1,\delta_2] &\iff& A(\gamma)\,,\\
\label{f13} \gamma\equiv [\delta_1,\delta_2] &\iff& S(\gamma) \an A(\gamma)\,.
\end{eqnarray}
\end{thm}

\proof
The proofs of (\ref{f11}) and (\ref{f12}) are straightforward.
Remember that for $p_1,p_2\in\om$, $(p_1,p_2)\mapsto \langle p_1,p_2\rangle $ and $\langle p_1,p_2\rangle\mapsto p_i$ ($i=1,2$) are computable.
(\ref{f13}) follows from  (\ref{f11}) and (\ref{f12}).
\qed

By (\ref{f11}), $[\delta_1,\delta_2]$ is (up to equivalence) the least, that is richest, multi-representation of $X_1\times X_2$ which can be ``synthesized'' from $\delta_1$ and $\delta_2$.
By (\ref{f12}), $[\delta_1,\delta_2]$ is (up to equivalence) the greatest, that is poorest, multi-representation of $X_1\times X_2$ which allows analysis, that is, allows to compute the components of a pair. In summary, $[\delta_1,\delta_2]$ is, up to equivalence, the only multi-representation that allows both, synthesis and analysis.
By Theorem~\ref{t4}, among the set of all  computability concepts on the Cartesian product, the one induced by the multi-representation $[\delta_1,\delta_2]$ is the most natural one.
The characterizations hold accordingly for finite and infinite products. Special cases have been considered, for example, in \cite[Lemma~3.3.4]{Wei00}.

We define the product of two, of finitely many and of countably many effective topological spaces as follows. The product of two effective topological spaces has been studied already in \cite[Section~8]{WG09}.

\begin{defi}[products of effective topological spaces]\label{d2} $ $\\
Let $\xx_i=(X_i,\tau_i,\beta_i,\nu_i)$, $i=1,2,\ldots$ be effective topological spaces.
\begin{enumerate}
 \item \label{d2a} Define the product $\xx_1\times \xx_2:=(X,\tau,\beta,\nu)$ as follows \cite{WG09} :\\
$X:=X_1\times X_2$, $\dom(\nu):=\{\langle u_1,u_2\rangle\mid u_1\in\dom(\nu_1),\ u_2\in\dom(\nu_2)\}$, $\nu\langle u_1,u_2\rangle:=\nu_1(u_1)\times \nu_2(u_2)$, $\beta:=\range(\nu)$, $\tau$ is the  topology generated by the set $\beta$.

\item \label{d2b}For  $n\geq 1$ define
$\xx_1\times \xx_2\times\ldots\times\xx_n:=\xx'_n:=(X'_n,\tau'_n,\beta'_n,\nu'_n)$ inductively by $\xx'_1:= \xx_1$, \ $\xx'_{n+1}:=\xx'_n\times \xx_{n+1}$, that is, $\xx_n'=(\ldots(\xx_1\times \xx_2)\times\ldots\times\xx_n)$.

\item \label{d2c} Define the countable product $\xx_1\times \xx_2\times\ldots:=\yy:=(Y,\tau_\yy,\beta_\yy,\nu_\yy)$ by
\begin{eqnarray*}
Y &:=& X_1\times X_2\times\ldots\,,\\
\dom(\nu_\yy)&:=&\{\langle u_1,\ldots,u_k\rangle\mid k\geq 1,\
u_i \in\dom(\nu_i) \ \mbox{for} \ 1\leq i\leq k\}\,,\\
\nu_\yy\langle u_1,\ldots,u_k\rangle&:=&\nu_1(u_1)\times\nu_2(u_2)\times\ldots\times\nu_k(u_k)\times  X_{k+1}\times X_{k+2}\times\ldots\,,\\
\beta _\yy &:=& \range (\nu_\yy)\,,\\
\tau_\yy & := & \mbox{the topology on $\yy$ generated by } \beta_\yy\,.
\end{eqnarray*}
\end{enumerate}
\end{defi}

In (\ref{d2a}), $\beta$ is a base of the product topology $\tau$ on $X_1\times X_2$,
In (\ref{d2b}), $\beta'_n$ is a base of the product topology $\tau'_n$ on $X_1\times\ldots\times X_n$, and in (\ref{d2c}), $\beta_\yy$ is a base of the product topology \,$\tau_\yy$ on $Y$ \cite{Eng89}. Therefore all the constructed spaces are effective topological spaces.
By the inductive definition in (\ref{d2b}),
\begin{eqnarray}\label{f3}
\nu'_n\langle\ldots\langle u_1,u_2\rangle,\ldots\rangle,u_n\rangle&=&\nu_1(u_1)\times\ldots\times \nu_n(u_n)
\end{eqnarray}
Notice that the following functions $c$ and $c'$ are computable (where $u_i\in\s$):
\begin{eqnarray}
\label{f1}c:1^n0\langle\ldots\langle u_1,u_2\rangle,\ldots\rangle,u_n\rangle &\mapsto& \langle u_1,\ldots,u_n\rangle \,,\\
\label{f2} c':\langle u_1,\ldots,u_n\rangle &\mapsto & \langle\ldots\langle u_1,u_2\rangle \ldots\,\rangle, u_n\rangle\,.
\end{eqnarray}

For each of the products of effective topological spaces  we have two representations of points which turn out to be equivalent.

\begin{thm}\label{t7} For effective topological spaces $\xx_1,\xx_2,\ldots$,
\begin{eqnarray}
\label{f14}[\delta_{\xx_1},\ldots,\delta_{\xx_n}]&\equiv &  \delta_{\xx_1\times \ldots\times \xx_n}\ \ \,,\\
\label{f15}[\delta_{\xx_1},\delta_{\xx_2}, \ldots\,]^+ &\equiv &
(\delta_{\xx'_1}\vee \delta_{\xx'_2}\vee \ldots)\,,\\
\label{f16}   [\delta_{\xx_1},\delta_{\xx_2},\ldots\,]&\equiv &
\delta_{\xx_1\times\xx_2\times\ldots}\,.
\end{eqnarray}
There are realizations of the reductions which do no depend on $\Delta$-names of the spaces $\xx_i$.
\end{thm}

(\ref{f14}) generalizes  \cite[Lemma~27]{WG09}. By
(\ref{f15}), $[\delta_{\xx_1}, \ldots,\delta_{\xx_n}]\equiv
\delta_{\xx_1\times\ldots\times \xx_n}$ uniformly in~$n$. \\

\proof

(\ref{f14}) Follows from (\ref{f15})
\medskip

(\ref{f15})$[\delta_{\xx_1},\delta_{\xx_2}, \ldots\,]^+(q)=(n,x_1,\ldots,x_n)$ iff
for some $p_1,\ldots, p_n$, $q=1^n0\langle p_1,\ldots, p_n\rangle $ and $p_i$ is a list of all $u_i$ such that $x_i\in\nu_i(u_i)$ ($i=1,\ldots,n$).
On the other hand, $(\delta_{\xx'_1}\vee \delta_{\xx'_2}\vee \ldots)(r)=(n,x_1,\ldots,x_n)$ iff for some $s$, $r=1^n0s$ and $s$ is a list of all $\langle\ldots\langle u_1,u_2\rangle,\ldots\rangle,u_n\rangle$ such that $x_i\in\nu_i(u_i)$ ($i=1,\ldots,n$).
Therefore, from $q$ we can find some $r$ such that $[\delta_{\xx_1},\delta_{\xx_2}, \ldots\,]^+(q)=(\delta_{\xx'_1}\vee \delta_{\xx'_2}\vee \ldots)(r)$ and vice versa.
\medskip

(\ref{f16}) $[\delta_{\xx_1},\delta_{\xx_2},\ldots\,](p)=(x_1,x_2,\ldots)$ iff there are $p_1,p_2,\ldots$ such that $p=\langle p_1,p_2,\ldots\,\rangle$
and for all $i$, $p_i$ is a list of all $u_i$ such that $x_i\in\nu_i(u_i)$.
On the other hand, $\delta_{\xx_1\times\xx_2\times\ldots\,}(q)=(x_1,x_2,\ldots\,)$
iff $q$ is a list of all $\langle u_1,u_2,\ldots, u_k\rangle$ such that
$(x_1,x_2,\ldots\,)\in \nu_1(u_1)\times\nu_2(u_2)\times\ldots\times\nu_k(u_k)\times  X_{k+1}\times X_{k+2}\times\ldots\,$.
Therefore, from $p$ we can find some $q$ such that
$[\delta_{\xx_1},\delta_{\xx_2},\ldots\,](p)=\delta_{\xx_1\times\xx_2\times\ldots\,}(q)$ and vice versa.

In both cases the computable functions operate only on names of the points and do not  require $\Delta$-names of the spaces $\xx_i$.
\qed

Theorem~\ref{t7} can be considered as a justification of the definition of the product space  $\xx_1\times\xx_2$.
The products on the class $\bf \ETS$ of effective topological spaces are computable.

\begin{thm}\label{t2}\hfill
\begin{enumerate}
\item \label{t2a}The function ${\rm PD_2}:\mathcal {T\times T\mapsto T}$,
$(\xx_1,\xx_2)\mapsto\xx_1\times\xx_2$
is $(\Delta,\Delta,\Delta)$-computable.

\item \label{t2b}The function ${\rm PD}^+:\biguplus_{n\geq 1}\mathcal T^n\to\mathcal T$,  $(n,\xx_1,\ldots, \xx_n)\mapsto \xx_n'$ is $([\Delta]^+,\Delta)$-computable.

\item \label{t2c}The function $\,{\rm PD}^\IN:\mathcal T ^\IN\to\mathcal T$,
$(\xx_1,\xx_2,\ldots)\mapsto \xx_1\times\xx_2\times\ldots $  is $([\Delta]^\IN,\Delta)$-computable.
\end{enumerate}
\end{thm}

\proof
 Let $\xx_i=(X_i,\tau_i,\beta_i,\nu_i)\in\Delta(t_i)$, $t_i=\langle r_i,s_i\rangle$ ($i\geq 1$). Then $r_i$~enumerates the graph of the characteristic function of $\dom(\nu_i)$
and $s_i$ enumerates a set $S_i\In (\dom(\nu_i))^3$ such that
$\nu_i(u)\cap\nu_i(v)=\bigcup\{\nu_i(w)\mid (u,v,w)\in S_i\}$.
\medskip

(\ref{t2a}) Let
 $S:=\{(\langle u_1,u_2\rangle,\langle v_1,v_2\rangle,\langle w_1,w_2\rangle)\mid
(u_1,v_1,w_1)\in S_1,\ (u_2,v_2,w_2)\in S_2\}$. A straightforward calculation shows

$\nu\langle u_1,u_2\rangle\cap \nu\langle v_1,v_2\rangle= \bigcup\{\nu\langle w_1,w_2\rangle\mid (\langle u_1,u_2\rangle,\langle v_1,v_2\rangle,\langle w_1,w_2\rangle)\in S\}$.\\
(An enumeration of the graph of) the characteristic function $r$ of $\dom(\nu)$ can be computed from $r_1$ and $r_2$ and an enumeration $s$ of $S$ can be computed from $s_1$ and $s_2$. Therefore, a word $\langle r,s\rangle$ can be computed which is a $\Delta$-name  of $\xx_1\times\xx_2$.
\medskip

(\ref{t2b}) By (\ref{t2a}) there is a computable function $h\pf\om\times\om\to\om$ such that $h(t_1,t_2)$ is a $\Delta$-name of $(\xx_1\times \xx_2)$ if $t_1$ is a $\Delta$-name of $\xx_1$ and $t_2$ is a $\Delta$-name of $\xx_2$.
There is a computable function $g\pf\om\to\om$ such that
$g(10t_1)=t_1$, \
$g(1^{n+1}0\langle t_1,\ldots,t_n,t_{n+1}\rangle)
=h(g(1^n0\langle t_1,\ldots,t_n\rangle),t_{n+1})$.
Then $g(10t_1)=t_1$ is a $[\Delta]^+$-name of $\xx_1'=\xx_1$.
Suppose by induction that $g(1^n0\langle t_1,\ldots,t_n\rangle)$ is a $\Delta$-name of $ \xx_n'$. Then $g(1^{n+1}0\langle t_1,\ldots,t_n,t_{n+1}\rangle)
=h(g(1^n0\langle t_1,\ldots,t_n\rangle),t_{n+1})$
is a $\Delta$-name of $\xx_n'\times\xx_{n+1}=\xx_{n+1}'$.
Therefore, $g$ is a
$([\Delta]^+,\Delta)$ realization of the function
$(\xx_1,\ldots, \xx_n)\mapsto \xx_1\times\ldots\times\xx_n$.
\medskip

(\ref{t2c}) Suppose $(\xx_1,\xx_2,...\,)\in[\Delta]^\IN(q)$. Consider $\yy$ from Definition~\ref{d2}. An enumeration $ r_\yy$ of the graph of the characteristic function of $\dom(\nu_\yy)$ can be computed from the characteristic functions of the $\dom(\nu_i)$  which can be computed from $q$.

First, for given $\langle u_1,\ldots ,u_m\rangle, \; \langle v_1,\ldots ,v_n\rangle \in \dom(\nu_\yy)$ we want to compute \\
$ \nu_\yy\langle u_1,\ldots ,u_m\rangle\cap  \nu_\yy\langle v_1,\ldots ,v_n\rangle$ as a union of base elements.

Assume $1\leq m\leq n$. Let $u_{m+1}:=v_{m+1}$, $\ldots, u_n:=v_n$.

By (\ref{t2b}) from $n$ and $q$  we can compute a $\Delta$-name $q_n$ of
$\xx'_n:=(X'_n,\tau'_n,\beta'_n,\nu'_n)$ and hence an enumeration of a set $S_n$ computing the intersection on $\beta'_n$  according to (\ref{f30}). By (\ref{f3}), (\ref{f1}) and   (\ref{f2}),
 \begin{eqnarray*}
&& \nu_\yy\langle u_1,\ldots ,u_m\rangle\cap  \nu_\yy\langle v_1,\ldots , v_n\rangle \\
&=&  \nu_\yy\langle u_1,\ldots ,u_n\rangle\cap  \nu_\yy\langle v_1,\ldots , v_n\rangle \\
&=&  (\nu_1(u_1)\times \ldots\times\nu_n(u_n)\cap \nu_1(v_1)\times \ldots\times\nu_n(v_n))\times X_{n+1}\times X_{n+2}\times\ldots\\
&=&(\nu'_n\circ c' \langle u_1,\ldots ,u_n\rangle\cap  \nu'_n\circ c' \langle v_1,\ldots ,v_n\rangle )\times X_{n+1}\times X_{n+2}\times\ldots\\
&=&\bigcup \{\nu'_n(w)\mid (c'\langle u_1,\ldots ,u_n\rangle ,c'\langle v_1,\ldots , v_n\rangle ,w)\in S_n\} \times X_{n+1}\times X_{n+2}\times\ldots\\
&=&\bigcup \{\nu'_n\circ c'\langle w_1,\ldots,w_n\rangle\mid \\
&&\qquad (c'\langle u_1,\ldots ,u_n\rangle ,c'\langle v_1,\ldots , v_n\rangle ,c'\langle w_1,\ldots,w_n\rangle)\in S_n\} \times X_{n+1}\times X_{n+2}\times\ldots\\
&=&\bigcup \{ \nu_\yy\langle w_1,\ldots,w_n\rangle \mid \\
&&\qquad (c'\langle u_1,\ldots ,u_n\rangle ,c'\langle v_1,\ldots , v_n\rangle ,c'\langle w_1,\ldots,w_n\rangle)\in S_n\}\,.
\end{eqnarray*}
Let
\begin{eqnarray*}S_{mn} &:=&  \{(\langle u_1,\ldots ,u_m\rangle, \langle v_1,\ldots , v_n\rangle , \langle w_1,\ldots,w_n\rangle)\mid \\
&& u_i,v_i,w_i\in\dom(\nu_i) \ \  \mbox{ and }\ \ (c'(\overline u),c'(\overline v),c'(\overline w))\in S_n\}
\end{eqnarray*}
where $\overline u:=\langle u_1,\ldots,u_m,v_{m+1},\ldots, v_n\rangle$,
$\overline v:=\langle v_1,\ldots,v_n\rangle$ and $\overline w:=\langle w_1,\ldots,w_n\rangle$.
Then
\begin{eqnarray*}
&&\nu_\yy\langle u_1,\ldots ,u_m\rangle\cap  \nu_\yy\langle v_1,\ldots , v_n\rangle \\
&=&\bigcup \nu_\yy\langle w_1,\ldots,w_n\rangle \mid
(\langle u_1,\ldots ,u_m\rangle, \langle v_1,\ldots , v_n\rangle , \langle w_1,\ldots,w_n\rangle)\in S_{mn}
\end{eqnarray*}
An enumeration of the set set $S_{mn}$ can be computed from $m,n$ and $S_n$, hence form $m,n$ and $q$. Correspondingly sets $S_{m,n}$ for $m>n$ can be computed from $m,n$ and $q$. Since for $x_i,y_i\in\s$,
$\langle x_1,\ldots ,x_m\rangle =\langle y_1,\ldots, y_n\rangle$ implies $m=n$ and $x_i=y_i$ for all $i$,
$S_{kl}\cap S_{mn}=\emptyset$ for $(k,l)\neq (m,n)$. Let $S:=\bigcup_{m,n\in\IN}S_{mn}$. Then
\[\nu_\yy(u)\cap\nu_\yy(v)=\bigcup\{ \nu_\yy(w)\mid (u,v,w)\in S\}\,.   \]
An enumeration of $S$ can be computed from $q$.
In summary, the function $(\xx_1,\xx_2,\ldots)\mapsto \xx_1\times\xx_2\times\ldots $  is $([\Delta]^\IN,\Delta)$-computable.
\qed

Next we study decomposition of products into their components.
Let ${\bf Y}:=(\emptyset,\{\emptyset\},\{\emptyset\},\nu_2)$ with $\dom(\nu_2)=\s$, which is a computable topological space. Then for every effective topological space
$\xx_1=(X_1,\tau_1,\beta_1,\nu_1)$ with $\dom(\nu)=\s$,
${\bf Z}:=\xx_1\times {\bf Y}=(\emptyset,\{\emptyset\},\{\emptyset\},\nu)\}$  with $\nu\langle u_1,u_2\rangle =\emptyset$ for all $u_1,u_2\in\s$.
Therefore, the function $(\xx_1,\xx_2)\mapsto \xx_1\times \xx_2$ is not injective, hence in general from $\xx_1\times \xx_2$ we cannot compute $\xx_1$ or $\xx_2$.

Let $\xx_1,\xx_2$ and $\xx$ be the spaces from Definition~\ref{d2}.\ref{d2a} and assume that $X_1$ and $X_2$ are not empty. There must be words $w_1,w_2\in\xx$ such that $\nu_1(w_1)\neq\emptyset$ and $\nu_2(w_2)\neq\emptyset$. Then
$\dom(\nu_1)=\{u_1\mid \langle u_1,w_2\rangle\in\dom(\nu)\}$ and for every $u_1\in\dom(\nu_1)$, $\nu_1(u_1)={\rm pr}_1\circ \nu\langle u_1, w_2\rangle$. Therefore, $\xx_1$ and (correspondingly) $\xx_2$ are determined uniquely by their product $\xx_1\times \xx_2$. However, we do not know whether decomposition of the product is computable for non-empty effective topological spaces. We prove computable decomposition for a somewhat smaller class of spaces.
Let
\begin{eqnarray*}
\ETS_1 &=& \{ (X,\tau,\beta ,\nu)\in\ETS\mid X\neq\emptyset\}\,,\\
\ETS_2 &=& \{ (X,\tau,\beta ,\nu)\in\ETS\mid  X\neq\emptyset\ \mbox{ and } (\forall U\in\beta)\, U\neq\emptyset\}\,.
\end{eqnarray*}
Then $\ETS_2\In \ETS_1\In \ETS$.

\begin{thm} \label{t8} For spaces from $\ETS_2$,
\begin{enumerate}
\item\label{t8a}
the functions $(\xx_1\times \xx_2)\mapsto \xx_i$ ($i=1,2$) are   $(\Delta,\Delta)$-computable,
\item \label{t8d}for  $i\leq n$,  the function
    $(i,(n,\xx_1\times\ldots\times\xx_n))\mapsto \xx_i$ is
    $(\nu_\IN,[\Delta]^+,\Delta)$-computable,
\item \label{t8b} the function
$\xx_1\times\xx_2\times \ldots\mapsto \xx_1$ is $(\Delta,\Delta)$-computable,
\item \label{t8c}
the function
$\xx_1\times\xx_2\times \ldots\mapsto \xx_2\times \xx_3\times\ldots$ is $(\Delta,\Delta)$-computable.
\end{enumerate}
In  (\ref{t8a}) for the case $i=1$, $\xx_1\in\ETS_1$ is sufficient, in (\ref{t8d}) $\xx_i\in\ETS_1$ is sufficient, in  (\ref{t8b}) $\xx_1\in\ETS_1$ is sufficient, and in (\ref{t8c}) $\xx_i\in{\mathcal T}_1$ for $i\geq 2$ is sufficient.

\end{thm}

\proof Consider the terminology from Definition\ref{d2}.
\medskip

(\ref{t8a})
We show that $(\xx_1\times \xx_2)\mapsto \xx_1$ is computable.
Let $\xx_1\times \xx_2\in\Delta\langle r,s\rangle$. Then $r$ enumerates the graph of the characteristic function $\chi$ of
$\dom(\nu)=\langle \dom(\nu_1),\dom(\nu_2)\rangle$. Since $\dom(\nu_1)\neq \emptyset$ and $\dom(\nu_2)\neq \emptyset$, from $r$ we can find words $t_0,t\in \s$ such that $\langle t_0,t\rangle \in\dom(\nu)$, hence $t_0\in\dom(\nu_1)$ and $t\in\dom(\nu_2)$. Since $u\in\dom(\nu_1)\iff \langle u,t\rangle \in\dom(\nu)$,
we can compute an enumeration $r_1$ of the graph of  the characteristic function of $\dom(\nu_1)$ and also an enumeration of $\dom(\nu_1)$.

The sequence $s$ enumerates a set $S$ of triples $(\langle u_1,u_2\rangle,\langle v_1,v_2\rangle,\langle w_1,w_2\rangle)$ for computing the intersection of base elements (\ref{f30}).
 %
 For $u_1,v_1\in\dom(\nu_1)$,
\begin{eqnarray*}
(\nu_1(u_1)\cap \nu_1(v_1))\times \nu_2(t)
&=& \nu\langle u_1,t\rangle \cap \nu\langle v_1,t\rangle\\
&=&  \bigcup \{\nu_1( w_1)\times\nu_2(w_2)\mid (\langle u_1,t\rangle,\langle v_1,t\rangle, \langle w_1,w_2\rangle)\in S\}\\
&=& \bigcup \{\nu_1( w_1)\times\nu_2(t)\mid (\exists w_2)\,(\langle u_1,t\rangle,\langle v_1,t\rangle,
\langle w_1,w_2\rangle)\in S\}\\
&=& \bigcup \{\nu_1( w_1)\mid (\exists w_2)(\langle u_1,t\rangle,\langle v_1,t\rangle,
\langle w_1,w_2\rangle)\in S\} \times\nu(t)
\end{eqnarray*}
The third  ``$=$'' holds since $\nu_2(w_2)\neq\emptyset $.
Since $\nu_2(t)\neq\emptyset$,
\[ \nu_1(u_1)\cap \nu_1(v_1)\ = \ \bigcup \{\nu( w_1)\mid (\exists w_2)(\langle u_1,t\rangle,\langle v_1,t\rangle, \langle w_1,w_2\rangle)\in S\}\,.\]
Let
$ S_1:=\{(u_1,v_1,w_1)\in(\dom(\nu_1))^3\mid (\exists w_2)(\langle u_1,t\rangle,\langle v_1,t\rangle, \langle w_1,w_2\rangle)\in S\}$.
Then\\
$\nu_1(u_1)\cap \nu_1(v_1)\ = \ \bigcup \{\nu( w_1)\mid (u_1,v_1,w_1)\in S_1\}$.
Since $\dom(\nu_1)$ and $S_1$ can be computed, a $\Delta$-name of $\xx_1$ can be computed from $\langle r,s\rangle$.

Notice that assuming $\xx_1\in\ETS_1$ is sufficient.
\medskip

(\ref{t8d}) Apply (\ref{t8a}) repeatedly, use a Turing machine on represented sets \cite{TW11b}. As an example we show how to compute $(2,(4,\xx_1\times\ldots\times\xx_4))\mapsto \xx_2$.
\[\xx_1\times\ldots\times\xx_4=\xx_3'\times \xx_4\mapsto \xx_3'=\xx_2'\times\xx_3\mapsto \xx_2'=\xx_1\times\xx_2\mapsto \xx_2\]
Notice that assuming $\xx_2\in\ETS_1$ is sufficient.

\medskip
(\ref{t8b}) Let $\yy=\xx_1\times \xx_2\ldots $  as in Definition~\ref{d2} and let
$\yy\in\Delta\langle r,s\rangle$. Then $r$ enumerates the graph of the characteristic function $\chi$ of $\dom(\nu_\yy)$.

Since $\dom(\nu_1)\neq \emptyset$ and $\dom(\nu_2)\neq \emptyset$, from $r$ we can find words $t_0,t\in \s$ such that $\langle t_0,t\rangle \in\dom(\nu_\yy)$, hence $t_0\in\dom(\nu_1)$ and $t\in\dom(\nu_2)$. Since $u\in\dom(\nu_1)\iff \langle u,t\rangle \in\dom(\nu_\yy)$,
we can compute an enumeration $r_1$ of the graph of  the characteristic function of $\dom(\nu_1)$ and also an enumeration of $\dom(\nu_1)$.
 For $u_1,v_1\in\dom(\nu_1)$,
\begin{eqnarray*}
&&(\nu_1(u_1)\cap \nu_1(v_1))\times \nu_2(t)\times X_3\times\ldots\\
&=& \nu_\yy\langle u_1,t\rangle \cap\nu_\yy\langle u_2, t \rangle\\
&=& \bigcup \{\nu_\yy\langle w_1,w_2,\ldots, w_n\rangle\mid (\langle u_1,t \rangle,
\langle u_2, t \rangle,\langle w_1,w_2,\ldots, w_n\rangle)\in S_\yy\}\\
&=& \bigcup \{\nu_\yy\langle w_1\rangle\mid (\langle u_1,t \rangle,
\langle u_2, t \rangle,\langle w_1\rangle)\in S_\yy\}\\
&& \cup\bigcup \{\nu_\yy\langle w_1,\ldots,w_n\rangle\mid  n\geq 2 \ \mbox{ and }\
\,(\langle u_1,t \rangle,
\langle u_2, t \rangle,\langle w_1,w_2,\ldots, w_n\rangle)\in S_\yy\}\\
&=& \bigcup \{\nu_\yy\langle w_1,t\rangle \mid (\langle u_1,t \rangle,
\langle u_2, t \rangle,\langle w_1\rangle)\in S_\yy\} \\
&& \cup \bigcup \{\nu_\yy\langle w_1,t\rangle\mid
  (\exists n\geq 2) \  (\exists w_2,\ldots w_n)\,(\langle u_1,t \rangle,
\langle u_2, t \rangle,\langle w_1,w_2,\ldots, w_n\rangle)\in S_\yy\}\\
&=&  \bigcup \{\nu_\yy\langle w_1,t\rangle\mid
(\exists n\geq 1) \   (\exists w_2,\ldots w_n)\,(\langle u_1,t \rangle,
\langle u_2, t \rangle,\langle w_1,w_2,\ldots, w_n\rangle)\in S_\yy\}\\
&=&\bigcup \{\nu_1(w_1)\times\nu_2(t)\times X_3\times \ldots\mid \\
&&\quad (\exists n\geq 1)\ (\exists w_2,\ldots w_n)\,(\langle u_1,t \rangle,
\langle u_2, t \rangle,\langle w_1,w_2,\ldots, w_n\rangle)\in S_\yy\}
\end{eqnarray*}
The fourth  ``$=$'' holds since $\nu_i(w_i)\neq\emptyset$ for all $i\geq 2$ and $X_i\neq\emptyset$ for all $i\geq 3$.
Since $\nu_2(t)\neq\emptyset$ and  $X_i\neq\emptyset$ for $i\geq 3$,
\[ \nu_1(u_1)\cap \nu_1(v_1)=
\bigcup \{\nu_1(w_1)\mid (\exists n\geq 1)\ (\exists w_2,\ldots w_n)\,(\langle u_1,t \rangle,
\langle u_2, t \rangle,\langle w_1,w_2,\ldots, w_n\rangle)\in S_\yy\}\,.\]
Let

 $S_1:=\{(u_1,v_1,w_1)\mid(\exists n\geq 1)\ (\exists w_2,\ldots w_n)\,(\langle u_1,t \rangle,
\langle u_2, t \rangle,\langle w_1,w_2,\ldots, w_n\rangle)\in S_\yy\}\,.$\\
Then $\nu_1(u_1)\cap\nu_1(v_1)=\bigcup \{\nu_1(w_1)\mid
(u_1,v_1,w_1)\in S_1\}$. From $s$, which enumerates $S_\yy$, we can compute an enumeration of $s_1\in\om$ of the set $S_1$. Therefore, we can compute a $\Delta$-name $\langle r_1,s_1\rangle$ of $\xx_1$.
Notice that assuming $\xx_1\in\ETS_1$ is sufficient.
\medskip

(\ref{t8c}) This proof is similar to those of (\ref{t8a}) and (\ref{t8b}).
\qed

By Theorems~\ref{t2} and~\ref{t8} many rearrangements of products of effective topological spaces from $\ETS_2$ are computable, for example\\
-- $\xx_1\times \xx_2\mapsto \xx_2\times\xx_1$,\\
-- $(\xx_1\times\xx_2)\times \xx_3 \mapsto \xx_1\times(\xx_2\times \xx_3)$,\\
-- $\xx\mapsto\xx\times\xx$, \ \ \ $\xx\mapsto \xx\times\xx\times\ldots$,\\
-- $\xx_1\times\yy_1\times\ldots\times\xx_n\times \yy_n\mapsto \xx_1\times\ldots\times\xx_n$, \ \ \\
-- $\xx_1\times \ldots\times\xx_n\mapsto \xx_n\times \ldots\times\xx_1$,\\
-- $(\xx_1\times \ldots\times\xx_m,\yy_1\times\ldots\times\yy_n)\mapsto
\xx_1\times \ldots\xx_m\times\yy_1\times\ldots\times \yy_n$ ),\\
-- $\xx_1\times\xx_2\times\ldots \mapsto \xx_{h(1)}\times \xx_{h(2}\times\ldots$ where $h:\IN\to\IN$ is computable.

We do not know whether Theorem~\ref{t8} remains valid for spaces from $\ETS_1$ where base elements may be empty. However,  rearrangements within products without deleting factors are possible on the whole space $\ETS$.

The product $(\xx_1,\xx_2)\mapsto \xx_1\times \xx_2$  is essentially commutative and associative.

\begin{thm}\label{t10}\hfill
\begin{enumerate}
\item \label{t10c}  The function $\xx_1\mapsto \xx_1\times \xx_1$ is $(\Delta,\Delta)$-computable.

\item \label{t10a} The function  $\xx_1\times \xx_2\mapsto\xx_2\times\xx_1$ is $(\Delta,\Delta)$-computable.
\item \label{t10b} The function $(\xx_1\times \xx_2)\times\xx_3\mapsto \xx_1\times(\xx_2\times \xx_3)$ is $(\Delta,\Delta)$-computable.
\item \label{t10d} The function $\xx_1\times(\xx_2\times \xx_3)\mapsto (\xx_1\times \xx_2)\times\xx_3 $ is $(\Delta,\Delta)$-computable.\end{enumerate}
\end{thm}

\proof For $i\geq 1$ let $\xx_i=(X_i,\tau_i,\beta_i,\nu_i)$.

(\ref{t10c}) This follows from computability of $(\xx_1,\xx_2)\mapsto \xx_1\times \xx_2$ (Theorem~\ref{t2}.\ref{t2a}).

(\ref{t10a})  Let $G \pf\ETS \to \ETS$ such that $\dom(H)=\{ \xx_1\times\xx_2\mid \xx_1,\xx_2\in\ETS\}$ and
$G( \xx_1\times\xx_2)=\xx_2\times\xx_1$.
Let $G(\yy)=\zz$. Then there are spaces $\xx_1,\xx_2$ such that
\begin{eqnarray*}
\yy& =& (Y,\tau_Y,\beta_Y,\nu_Y)=\xx_1\times\xx_2\,,\\
\zz & = & (Z,\tau_Z,\beta_Z,\nu_Z)=\xx_2\times \xx_1\,.
\end{eqnarray*}
where $Y=X_1\times X_2$ and $Z=X_2\times X_1$,
\begin{eqnarray*}
\dom(\nu_Y) &=& \{ \langle u_1, u_2\rangle\mid (\forall 1\leq i\leq 2)\,u_i\in\dom(\nu_i)\}\,,\\
\dom(\nu_Z) &=& \{\langle v_2, v_1\rangle\mid (\forall 1\leq i\leq 2)\,v_i\in\dom(\nu_i)\}\,.
\end{eqnarray*}
Suppose $\Delta\langle p,q\rangle=\yy$. Then the sequence $p$ enumerates the graph of the characteristic function of $\dom(\nu_Y)$ and the sequence $q$ enumerates some set $S_\yy$ such that $\nu_\yy(u)\cap\nu_\yy(v)=\bigcup \nu_\yy(w)\mid \langle u,v,w\rangle\in S_\yy$. Since $\langle u_1,u_2\rangle\in\dom(\nu_\yy)\iff \langle u_2,u_1\rangle\in\dom(\nu_\zz)$, from $p$ we can compute some $s\in\om$ which enumerates $\dom(\nu_\zz)$.
Define $S_\zz\In (\dom(\nu_\zz))^3$ by
\[ S_\zz :=
\{(\langle u_2, u_1\rangle,\langle v_2, v_1\rangle,\langle w_2, w_1\rangle)
\mid
(\langle u_1, u_2\rangle ,  \langle v_1,v_2\rangle, \langle w_1, w_2\rangle)\in S_\yy\}\,.
\]
Then
\begin{eqnarray*}
&&\nu_Z\langle u_2,u_1\rangle  \cap
\nu_Z\langle v_2,v_1\rangle\\
&=& \nu_2(u_2)\times\nu_1(u_1)\cap
\nu_2(v_2)\times\nu_1(v_1)\\
&=& \bigcup \{\nu_2(w_2)\times\nu_1(w_1)\\
&&\quad\mid (\langle u_1, u_2\rangle ,  \langle v_1,v_2\rangle, \langle w_1, w_2\rangle)\in S_\yy\}\\
&=& \bigcup \{\nu_\zz\langle w_2,w_1\rangle\\
&&\quad\mid (\langle u_2, u_1\rangle ,  \langle v_2,v_1\rangle, \langle w_2, w_1\rangle)\in S_\zz\}
\end{eqnarray*}
From the enumeration $q$ of $S_\yy$ we can compute an enumeration $s$ of $S_\zz$.
Therefore, $\Delta\langle r,s\rangle=\zz$ and hence the operator $G$ is computable.

(\ref{t10b}) Let $H \pf\ETS \to \ETS$ such that $\dom(H)=\{ (\xx_1\times\xx_2)\times \xx_3\mid \xx_1,\xx_2,\xx_3\in\ETS\}$ and
$H( (\xx_1\times\xx_2)\times\xx_3)=\xx_1\times (\xx_2\times\xx_3)$.
Let $H(\yy)=\zz$. Then there are spaces $\xx_1,\xx_2,\xx_3$ such that
\begin{eqnarray*}
\yy& =& (Y,\tau_Y,\beta_Y,\nu_Y)=(\xx_1\times\xx_2)\times\xx_3\,,\\
\zz & = & (Z,\tau_Z,\beta_Z,\nu_Z)=\xx_1\times(\xx_2\times \xx_3)\,.
\end{eqnarray*}
By Definition~\ref{d2}, $Y=Z=X_1\times X_2\times X_3$ and $\tau_\yy=\tau_\zz$,
\begin{eqnarray*}
\dom(\nu_Y) &=& \{\langle \langle u_1, u_2\rangle, u_3\rangle\mid (\forall 1\leq i\leq 3)\,u_i\in\dom(\nu_i)\}\,,\\
\dom(\nu_Z) &=& \{\langle u_1,\langle u_2, u_3\rangle\rangle\mid (\forall 1\leq i\leq 3)\,u_i\in\dom(\nu_i)\}\,.
\end{eqnarray*}
Suppose $\Delta\langle p,q\rangle=\yy$. Then the sequence $p$ enumerates the graph of the characteristic function of $\dom(\nu_Y)$ and the sequence $q$ enumerates some set $S_\yy$ such that $\nu_\yy(u)\cap\nu_\yy(v)=\bigcup \{ \nu_\yy(w)\mid \langle u,v,w\rangle\in S_\yy\}$.

Therefore, from $p$ we can compute an enumeration $r$ of the graph of the characteristic function of $\dom(\nu_Z)$. Define $S_\zz\In (\dom(\nu_\zz))^3$
by
\begin{eqnarray*}
S_\zz &:=&
\{(\langle u_1,\langle u_2, u_3\rangle\rangle,\langle v_1,\langle v_2, v_3\rangle\rangle,\langle w_1,\langle w_2, w_3\rangle\rangle)\\
&& \qquad \mid
(\langle \langle u_1, u_2\rangle ,u_3\rangle,
\langle \langle v_1,v_2\rangle, v_3\rangle,
\langle \langle w_1, w_2\rangle ,w_3\rangle)\in S_\yy\}\,.
\end{eqnarray*}
Then
\begin{eqnarray*}
&&\nu_Z\langle u_1,\langle u_2,u_3\rangle\rangle  \cap
\nu_Z\langle v_1,\langle v_2,v_3\rangle\rangle\\
&=& \nu_1(u_1)\times\nu_2(u_2)\times\nu_3(u_3)\cap
\nu_1(v_1)\times\nu_2(v_2)\times\nu_3(v_3)\\
&=& \nu_Y\langle \langle u_1, u_2\rangle,u_3\rangle\cap
\nu_Y\langle \langle v_1, v_2\rangle,v_3\rangle\\
&=& \bigcup\{\nu_\yy\langle \langle w_1, w_2\rangle,w_3\rangle\rangle\mid \\
&& \hspace{5ex}
(\langle \langle u_1, u_2\rangle,u_3\rangle,  \langle \langle v_1, v_2\rangle,v_3\rangle,   \langle \langle w_1, w_2\rangle,w_3\rangle)\in S_\yy\}\\ &=& \bigcup\{\nu_\zz\langle  w_1, \langle w_2,w_3\rangle\rangle\mid \\
&& \hspace{5ex}(\langle u_1,\langle u_2, u_3\rangle\rangle,\langle v_1,\langle v_2, v_3\rangle\rangle,\langle w_1,\langle w_2, w_3\rangle\rangle)\in S_\zz\}
\end{eqnarray*}
From the enumeration $q$ of $S_\yy$ we can compute an enumeration $s$ of $S_\zz$.
Therefore, $\Delta\langle r,s\rangle=\zz$ and hence the operator $H$ is computable.

(\ref{t10d}) Analog to (\ref{t10b}).
\qed

These results can be generalized to longer products. We do not go into further details.

\section{Products of subsets, Tychonoff's theorem}\label{sece}

In this section we prove that the various product operations are computable on points, on arbitrary sets and on compact sets.
Some facts about the product of two computable topological spaces are already proved in \cite[Lemma~27]{WG09} (where the proof of Lemma~27.7 on the product of compact sets is false).  Here we prove uniform versions also for finite and for infinite products.
By (\ref{f9}), the multi-representation $\delta$ of points can be considered as the restriction of the multi-representation $\tp$ to the sets with cardinality~$1$.
Therefore, we start with the multi-representation~$\tp$ of sets.
Remember that $(\xx,B)\in\tp(p):\iff B\in\tp_\xx(p)$.

\begin{thm}\label{t5}\hfill
\begin {enumerate}
\item\label{t5a} The function \,$\Pi^{s2}:(\ETSS)^2 \to  \ETSS$,\\
  \hspace*{5ex} $((\xx_1,B_1),(\xx_2,B_2))\mapsto (\xx_1\times \xx_2,B_1\times B_2)$, \\
   is $([\tp,\tp],\tp))$-computable.
\item\label{t5b} The function \,$\Pi^{s+}:\biguplus_{n\geq 1}(\ETSS)^n \to \ETSS$,\\
   \hspace*{5ex}   $(n,(\xx_1,B_1),\ldots,(\xx_n,B_n))\mapsto (\xx_1\times \ldots\times \xx_n,B_1\times\ldots\times B_n)$,\\
  is $([\tp]^+,\tp)$-computable.
\item \label{t5c}
 The function \,$\Pi^{s\infty}:(\ETSS)^\IN \to  \ETSS$, \\
 \hspace*{5ex} $((\xx_1,B_1),(\xx_2,B_2),\ldots) \mapsto(\xx_1\times\xx_2\times \ldots,B_1\times B_2\times\ldots)$,\\  is $([\tp]^\IN,\tp)$-computable.
\end{enumerate}
The three functions are also computable w.r.t.  $\,\tp^\Delta$ instead of $\,\tp$.
\end{thm}

\proof

(\ref{t5a}) Let $M$ be a Type-2 Turing machine that on input $\langle p_1,p_2\rangle$ writes a sequence of all $\iota\langle u_1,u_2\rangle$ such that $u_1\ll p_1$ and $u_2\ll p_2$ (and from time to time writes $11$ in order to produce an infinite sequence).
Since
\[\begin{array}{llll}
&((\xx_1,B_1),(\xx_2,B_2))\in [\tp,\tp]\langle p_1,p_2\rangle&\Longrightarrow & B_1\in\tp_{\xx_1}(p_1) $ and $B_2\in\tp_{\xx_2}(p_2)\\
\Longrightarrow &B_1\times B_2\in \tp_{\xx_1\times\xx_2}\circ f_M\langle p_1,p_2\rangle &\Longrightarrow &
(\xx_1\times\xx_2,B_1\times B_2)\in\tp\circ f_M\langle p_1,p_2\rangle
\end{array}\]
(notice that $\nu_1(u_1)\cap B_1\neq\emptyset$ and $\nu_2(u_2)\cap B_2\neq\emptyset$ iff $\nu\langle u_1,u_2\rangle\cap B_1\times B_2\neq\emptyset$),
the computable function $f_M$ is a realizer of $\Pi^{s2}$.
\medskip

(\ref{t5b}) Let $f_M$ be the computable realization from Case (\ref{t5a}).
There is a computable function $H :\om\to\om$ such that
\begin{eqnarray*}
H(10\langle p_1\rangle) &=& p_1\,,\\
H (1^{n+1}0\langle p_1,\ldots,p_{n+1}\rangle)
&=&f_M (H ( 1^n0\langle p_1,\ldots,p_{n}\rangle), p_{n+1})\,.
\end{eqnarray*}
We show by induction that the function $H$  realizes the function $\Pi^{s+}$. For $n=1$ we obtain:
\[
(1,(\xx_1,B_1))\in[\tp]^+(10\langle p_1\rangle)
\Longrightarrow (\xx_1,B_1)\in\tp (p_1)
\Longrightarrow \Pi^{s+}(1,(\xx_1,B_1))\in \tp \circ H(10\langle p_1\rangle)\,.\]
and for $n+1$ by induction,
\begin{eqnarray*}
&&(n+1,(\xx_1,B_1),\ldots,(\xx_{n+1},B_{n+1}))\in [\tp]^+( 1^{n+1}0\langle p_1,\ldots,p_{n+1}\rangle)\\
&\Longrightarrow & (\xx_i,B_i)\in \tp (p_i)\ \ \mbox{for}\ \ 1\leq i\leq n+1\\
&\Longrightarrow & ((\xx_1,B_1),\ldots,(\xx_n,B_n))\in [\tp]^+(1^n0\langle p_1,\ldots,p_n\rangle) \an
(\xx_{n+1},B_{n+1})\in \tp (p_{n+1})\\
&\Longrightarrow & (\xx_1\times \ldots\times \xx_n,B_1\times \ldots\times B_n)\in\tp\circ H(1^n0\langle p_1,\ldots,p_{n}\rangle)\\
&&\qquad \an (\xx_{n+1},B_{n+1})\in \tp (p_{n+1})\\
&\Longrightarrow & (\xx_1\times\ldots\times\xx_{n+1},B_1\times\ldots\times B_{n+1})\in \tp\circ
f_M( H(1^n0\langle p_1,\ldots,p_{n}\rangle), p_{n+1})\\
&\Longrightarrow & \Pi^{s+}(n+1,(\xx_1,B_1),\ldots,(\xx_{n+1},B_{n+1})\in \tp\circ H(1^{n+1}0\langle p_1,\ldots,p_{n}, p_{n+1}\rangle)\\\,.
\end{eqnarray*}

(\ref{t5c}) Suppose
$((\xx_1,B_1),(\xx_2,B_2),\ldots)\in[\tp]^\IN\langle p_1,p_2,\ldots\rangle$.
Then $(\forall i)(\xx_i,B_i)\in\tp(p_i)$. Therefore, for every $i$, $p_i$ is a list of all $\iota(u_i)$ such that ($u_i\in\dom(\nu_i)$ and) $\nu_i(u_i)\cap B_i\neq\emptyset$.
With the terminology of Definition~\ref{d2}(\ref{d2c}), from $\langle p_1,p_2,\ldots\rangle$
we want to compute a list of all $\iota(w)$ such that $w\in \dom(\nu_\yy)$ and $B_1\times B_2\times \ldots\cap \nu_\yy(w)\neq\emptyset$.
By the definition,  $w\in\dom(\nu_\yy)$ iff there are some $k\geq 1$ and words
$u_1\in\dom(\nu_1)$, ... , $u_k\in\dom(\nu_k)$ such that
$w=\langle u_1u_2\ldots u_k\rangle=\iota(u_1)\iota(u_2)\ldots\iota(u_k)$. For every $k\geq 1$,
\begin{eqnarray*}
&&\nu_\yy(\iota(u_1),\ldots,\iota(u_k))\cap B_1\times B_2\times\ldots\neq\emptyset\\
&\iff & (\forall 1\leq i\leq k)\, \nu(u_i)\cap B_i\neq\emptyset\\
&\iff & (\forall 1\leq i\leq k)\, u_i\ll p_i
\end{eqnarray*}
There is a computable function $h$ such that $h\langle p_1,p_2,\ldots\rangle$ is a list of all $\iota(\iota(u_1)\iota(u_2)\ldots\iota(u_k))$ such that $k\geq 1$ and $u_i\ll p_i$ for all $1\leq i\leq k$. Then
\[((\xx_1,B_1),(\xx_2,B_2),\ldots)\in[\tp]^\IN\langle p_1,p_2,\ldots\rangle
\Longrightarrow (Y,B_1\times B_2\times\ldots)\in \tp\circ h\langle p_1,p_2,\ldots\rangle\,.\]
Therefore, $\Pi^{s\infty}$ is $([\tp]^\IN,\tp)$-computable.
\medskip

By Theorem~\ref{t2} the three functions are also computable w.r.t.  $\,\tp^\Delta$ instead of $\,\tp$.
\qed

In the proof for the case of $\tp$ for every effective topological space $\xx=(X,\tau,\beta,\nu)$, no information about intersections $\nu(u)\cap\nu(v)$ is needed and the information about $\dom(\nu)$ given by the $\tp$-names is sufficient. Therefore the computable realizations of the operators are independent of the spaces, hence the theorem can be formulated for $\tp$.

The next theorem considers points.
\newpage

\begin{thm}\label{t3}\hfill
\begin {enumerate}
\item\label{t3a} The function \,$\Pi^{p2}:(\ETSP)^2 \to  \ETSP$,\\
 \hspace*{5ex} $((\xx_1,x_1),(\xx_2,x_2))\mapsto (\xx_1\times \xx_2,(x_1,x_2))$, \\is $([\delta,\delta],\delta))$-computable.
    Its inverse is $(\delta,[\delta,\delta])$-computable.
\item\label{t3b} The function \,$\Pi^{p+}:\biguplus_{n\geq 1}(\ETSP)^n \to \IN\times \ETSP$  ,\\
     \hspace*{5ex} $(n,(\xx_1,x_1),\ldots,(\xx_n,x_n))\mapsto(n, (\xx_1\times \ldots\times \xx_n,(x_1,\ldots,x_n)))$,\\
is $([\delta]^+,[\nu_\IN,\delta])$-computable.
Its inverse is $([\nu_\IN,\delta],[\delta]^+)$-computable.

\item \label{t3c}
The function \,$\Pi^{p\infty}:(\ETSP)^\IN \to  \ETSP$, \\
 \hspace*{5ex} $((\xx_1,x_1),(\xx_2,x_2),\ldots)\mapsto(\xx_1\times\xx_2\times \ldots,(x_1,x_2,\ldots))$,\\
  is $([\delta]^\IN,\delta)$-computable. Its inverse is  $(\delta,[\delta]^\IN)$-computable.
\end{enumerate}
The three functions are also computable w.r.t. $\delta^\Delta$ instead of $\delta$. Their inverses restricted to $\ETS _2$ (the spaces with non-empty base sets) are computable w.r.t. $\delta^\Delta$ instead of $\delta$.
\end{thm}

This theorem is an other formulation of Theorem~\ref{t7}.
By (\ref{f9}), computability of $\Pi^{p2}$, $\Pi^{p+}$ and $\Pi^{p\infty}$ follows from Theorem~\ref{t5}. \\

\proof
By  (\ref{f9}) and Theorem~\ref{t5} the functions $\Pi^{p2}$, $\Pi^{p+}$ and $\Pi^{p\infty}$ can be considered as restrictions of the computable functions
$\Pi^{s2}$, $\Pi^{s+}$ and $\Pi^{s\infty}$ to singleton sets, hence they are computable.
It remains to show that their inverses are computable.
\medskip

(\ref{t3a}) Suppose $(\xx_1\times\xx_2,(x_1,x_2))\in\delta(p)$. Then $(x_1,x_2)=\delta_{\xx_1\times\xx_2}(p)$. By Theorem~\ref{t7} there is a computable function $h$ such that $(x_1,x_2)=[\delta_{\xx_1},\delta_{\xx_2}]\circ h(p)$. There are computable functions $h_1,h_2$ such that $h(p)=\langle h_1(p),h_2(p)\rangle$. Then $(x_1,x_2)=(\delta_{\xx_1}\circ h_1(p),\delta_{\xx_2}\circ h_2(p))$, hence $(\xx_1,x_1)\in\delta\circ h_1(p)$ and $(\xx_2,x_2)\in\delta\circ h_2(p)$ and finally
$((\xx_1,x_1),(\xx_2,x_2))\in[\delta,\delta]\langle h_1(p),h_2(p) =
[\delta,\delta]\circ h(p)$.
\medskip

(\ref{t3b}) Suppose $(n, (\xx_1\times \ldots\times \xx_n,(x_1,\ldots,x_n)))\in[\nu_\IN,\delta]\langle p,q\rangle$.
Then $\nu_\IN(p)=n$ and $(x_1,\ldots,x_n)= \delta_{\xx_1\times\ldots\times\xx_n}(q)$, hence
$(n, x_1,\ldots, x_n)=(\delta_{\xx_1'}\vee\delta_{\xx_2'}\vee\ldots)(1^n0q)$. By Theorem~\ref{t7} there is a computable function $h$ such that
$(n, x_1,\ldots, x_n)=[\delta_{\xx_1},\delta_{\xx_2},\ldots]^+\circ h(1^n0q)$. There are functions $p_1,\ldots,p_n\in\om$ such that $h(1^n0q)=1^n0\langle p_1,\ldots,p_n\rangle$. Then by Definition~\ref{d3}.\ref{d3c}, $x_i=\delta_{\xx_i}(p_i)$, hence $(\xx_i,x_i)\in\delta(p_i)$ for all $1\leq i\leq n$. Hence, $(n,(\xx_1,x_1),\ldots, (\xx_n,x_n))\in [\delta]^+(
1^n0\langle p_1,\ldots,p_n\rangle) = h(1^n0q)$. Therefore, $\langle p,q\rangle \mapsto h(1^{\nu_\IN(p)}0q)$ is a computable realization of $(\Pi^{p+})^{-1}$.
\medskip

\ref{t3c}) Suppose $(\xx_1\times\xx_2\times \ldots,(x_1,x_2,\ldots))\in\delta(p)$. Then $(x_1,x_2,\ldots)=\delta_{\xx_1\times\xx_2\times \ldots}(p)$. By Theorem~\ref{t7} there is a computable function $h$ such that
$\delta_{\xx_1\times\xx_2\times \ldots}(p)=[\delta_{\xx_1},\delta_{\xx_2},\ldots]\circ h(p)$. There are unique functions $p_i\in\om$ such that $h(p)=\langle p_1,p_2,\ldots\rangle$. Then $x_i=\delta_{\xx_i}(p_p)$, hence $(\xx_i,x_i)\in\delta(p_i)$ for all~$i$. Therefore, $((\xx_1,x_1),(\xx_2,x_2),\ldots)\in[\delta]^\IN\circ h(p)$.
\medskip

Computability w.r.t. $\delta^\Delta$ instead of $\delta$ follows from Theorems~\ref{t2} and~\ref{t8}.
\qed

Notice that a  name $p$ of $(\xx,x)=\delta(p)$ contains no information about the effective topological space  $\xx$. Therefore, the theorem does not  mean that the components $\xx_i$ of the product spaces can be computed in general (cf. Theorem~\ref{t8}).

\begin{cor}[projections]\label{c8} $ $

\begin{enumerate}
\item The functions $(\xx_1\times \xx_2,(x_1,x_2))\mapsto (\xx_i,x_i)$ ($i=1,2$) are $(\delta,\delta)$-computable.
\item The function

$(i,n, (\xx_1\times \ldots\times \xx_n,(x_1,\ldots,x_n)))\mapsto (\xx_i,x_i)$
($1\leq  i\leq n$) \\
is $(\nu_\IN,\nu_\IN,\delta,\delta)$-computable.
\item The function $(i,(\xx_1\times\xx_2\times\ldots,(x_1,x_2,\ldots)))\mapsto(\xx_i,x_i)$ is $(\nu_\IN,\delta,\delta)$-computable.
\end{enumerate}
\end{cor}

As main results we prove computable versions of Tychonoff's theorem stating that the product of compact spaces is compact.

\begin{thm}[computable Tychonoff theorem]\label{t6}\hfill
\begin {enumerate}
\item\label{t6a} The function \,$\Pi^{c2}:(\ETSC)^2 \to  \ETSC$,\\
 \hspace*{5ex} $((\xx_1,K_1),(\xx_2,K_2))\mapsto (\xx_1\times \xx_2,K_1\times K_2)$ \\
is $([{\kappa^\Delta},{\kappa^\Delta}],{\kappa^\Delta}))$-computable.

\item\label{t6b} The function \,$\Pi^{c+}:\biguplus_{n\geq 1}(\ETSC)^n \to \ETSC$  ,\\
     \hspace*{5ex} $(n,(\xx_1,K_1),\ldots,(\xx_n,K_n))\mapsto (\xx_1\times \ldots\times \xx_n,K_1\times\ldots\times K_n)$,\\
is $([{\kappa^\Delta}]^+,{\kappa^\Delta})$-computable.

 \item \label{t6c}
The function \,$\Pi^{c\infty}:(\ETSC)^\IN \to  \ETSC$, \\
 \hspace*{5ex} $((\xx_1,K_1),(\xx_2,K_2),\ldots)\mapsto(\xx_1\times\xx_2\times \ldots,K_1\times K_2\times\ldots)$ \\ is $([\kappa^\Delta]^\IN,{\kappa^\Delta})$-computable.
\end{enumerate}
\end{thm}

\proof  We use the terminology from Definition~\ref{d2}.

 (\ref{t6a}) For compact sets $K_1\In X_1$ and $K_2\In X_2$, $K_1\times K_2$ is
compact in $\xx_1\times \xx_2$ \cite{Eng89}.
First, we assume that $\xx_1$ and $\xx_2$ are {\em computable} topological spaces.
We want to enumerate all words $w$ such that $K_1\times K_2\In \bigcup\nufs(w)$.
From $w\in\dom(\nufs)$ we can compute a number $n$ and pairs $(u_i,v_i)\in \dom(\nu_1)\times \dom(\nu_2)$ ($1\leq i\leq n$) such that
$\bigcup\nufs(w)=\bigcup_{i\in I}\nu_1(u_i)\times \nu_2(v_i)$  for the index set $I:=\{1,\ldots,n\}$. Then
\begin{eqnarray}
&&K_1\times K_2 \In \bigcup\nufs(w)\\
&\iff &K_1\times K_2 \In \bigcup_{i\in I}\nu_1(u_i)\times \nu_2(v_i)\\
&\iff &
(\forall x\in K_1)(\exists J\In I)\, \Big(x\in \bigcap_{j\in J}\nu_1(u_j)\wedge
K_2\In \bigcup _{j\in J}\nu_2(v_j)\Big )\\
&\iff &(\exists J_1,\ldots,J_m\In I)\Big (K_1\In \bigcup_{l=1}^m\bigcap_{j\in J_l}\nu_1(u_j) \wedge (\forall l)\,K_2\In \bigcup _{j\in J_l}\nu_2(v_j)\Big)\\
&\iff &(\exists J_1,\ldots,J_m\In I)\Big (K_1\In \bigcup_{l=1}^m\bigcap_{j\in J_l}\nu_1(u_j) \wedge K_2\In \bigcap_{l=1}^m\bigcup _{j\in J_l}\nu_2(v_j)\Big)\,.
\end{eqnarray}

By \cite[Theorem~11]{WG09}, from $w$ and $J_1,\ldots, J_m$  (a $\theta_{\xx_1}$-name of) the open set $ W_1:=\bigcup_{l=1}^m\bigcap_{j\in J_l}\nu_1(u_j)$ and (a $\theta_{\xx_2}$-name of) the open set $W_2:=\bigcap_{l=1}^m\bigcup _{j\in J_l}\nu_2(v_j)$ can be computed.
By \cite[Theorem~13.5]{WG09}, $K\In W$ for compact $K$ and open $W$ is $(\kappa_{\xx_i},\theta_{\xx_i})$-r.e. for $i=1,2$.
Therefore, $K_1\times K_2\In  V$ is $(\kappa_{\xx_1},\kappa_{\xx_2},\bigcup\nufs)$-r.e. , hence
from (a $\kappa_{\xx_1}$-name of) $K_1$  and (a $\kappa_{\xx_2}$-name of) $K_2$ we can compute a list of all $w$ such that $K_1\times K_2 \In \bigcup\nufs(w)$. Therefore, there is a computable function $h\pf\om\times\om\to\om$ such that
\begin{eqnarray}\label{f8}\mbox{$K_1\times K_2\in \kappa_{\xx_1\times \xx_2}\circ h(p_1,p_2)$ if $K_1\in \kappa_{\xx_1}(p_1)$ and  $K_2\in \kappa_{\xx_2}(p_2)$}\,,
\end{eqnarray}
 hence
$(K_1,K_2)\mapsto K_1\times K_2$ is $(\kappa_{\xx_1},\kappa_{\xx_2},\kappa_{\xx_1\times \xx_2})$-computable.

If we abbreviate (\ref{f8}) by  $Q(h,\xx_1,\xx_2)$,
then by the meta-theorem~\ref{t1}, there is a computable function $\overline h$ such that
\begin{eqnarray*}
&K_1\times K_2  \in \kappa_{\xx_1\times \xx_2}\circ \overline h (t_1,t_2,p_1,p_2)\\
&\mbox{if
$\xx_1\in\Delta(t_1)$, $\xx_2\in\Delta(t_2)$, $K_1\in\kappa_{\xx_1}(p_1)$ and $K_2\in\kappa_{\xx_2}(p_2)$}
\end{eqnarray*}
By Theorem~\ref{t2} there is a computable function $f$ such that\\
\hspace*{10ex}$\xx_1\times\xx_2\in\Delta\circ f(t_1,t_2)$ \ if \ $\xx_1\in\Delta(t_1)$ and $\xx_2\in\Delta(t_2)$.\\
Therefore,
\begin{eqnarray*}
&(\xx_1\times\xx_2,K_1\times K_2)  \in \kappa^\Delta\langle f(t_1,t_2),\overline h (t_1,t_2,p_1,p_2)\rangle\\
&\mbox{if
$(\xx_1,K_1)\in\kappa^\Delta\langle t_1,p_1\rangle $ and
$(\xx_2,K_2)\in\kappa^\Delta\langle t_2,p_2\rangle $}
\end{eqnarray*}
Define $h'\langle\langle t_1,p_1\rangle,\langle t_2,p_2\rangle\rangle:=\langle f(t_1,t_2),\overline h (t_1,t_2,p_1,p_2)\rangle$. Then
\begin{eqnarray*}
(\xx_1\times\xx_2,K_1\times K_2)  \in \kappa^\Delta \circ h'\langle q_1,q_2\rangle
&\mbox{if} & \mbox{
$(\xx_1,K_1)\in\kappa^\Delta(q_1) $ and
$(\xx_2,K_2)\in\kappa^\Delta(q_2) $}
\end{eqnarray*}
Therefore,  the function $((\xx_1,K_1),(\xx_2,K_2))\mapsto (\xx_1\times\xx_2,K_1\times K_2)$  is $(\kappa^\Delta,\kappa^\Delta,\kappa^\Delta)$\bb computable.

(\ref{t6b})
Let $h'$ be the realizing computable function from Case (\ref{t6a}). There is a computable function $H :\om\to\om$ such that
\begin{eqnarray*}
H( 10\langle p_1\rangle )&=& p_1\,,\\
H(1^{n+1}0\langle p_1,\ldots,p_{n+1}\rangle)
&=& h'( H(1^n0\langle p_1,\ldots,p_{n}\rangle), p_{n+1})\,.
\end{eqnarray*}
We show by induction that the (computable) function $H$  realizes the function $\Pi^{c+}$. For $n=1$ we obtain:
\begin{eqnarray*}(1,(\xx_1,K_1))\in[\kappa^\Delta]^+( 10\langle p_1\rangle) &\Longrightarrow& (\xx_1,K_1)\in\kappa^\Delta(p_1)=\kappa^\Delta\circ H(10\langle p_1\rangle)\,.
\end{eqnarray*}
Suppose $(n+1,(\xx_1,K_1),\ldots, (\xx_{n+1},K_{n+1})\in[\kappa^\Delta]^+(p)$. Then there are $p_i\in\om$ such that
$p= 1^{n+1}0 \langle p_1,\ldots,p_{n+1}\rangle$. We obtain

\begin{eqnarray*}
&& (n+1,(\xx_1,K_1),\ldots, (\xx_{n+1},K_{n+1}))\in[\kappa^\Delta]^+
( 1^{n+1}0 \langle p_1,\ldots,p_{n+1}\rangle)\\
&\Longrightarrow &
(n,(\xx_1,K_1),\ldots, (\xx_{n+1},K_n)\in[\kappa^\Delta]^+\langle 1^n0
\langle p_1,\ldots,p_n\rangle\rangle \an (\xx_{n+1},K_{n+1})\in \kappa^\Delta(p_{n+1})\\
&\Longrightarrow & (\xx_1\times\ldots\times \xx_n, K_1\times\ldots\times K_n)\in
\kappa^\Delta\circ H ( 1^n0 \langle p_1,\ldots,p_n\rangle)\\
&&\qquad\an (\xx_{n+1},K_{n+1})\in \kappa^\Delta(p_{n+1})\\
&\Longrightarrow & (\xx_1\times\ldots\times \xx_{n+1}, K_1\times\ldots\times K_{n+1})\in
\kappa^\Delta\circ h'( H ( 1^n0 \langle p_1,\ldots,p_n\rangle),p_{n+1})\\
&\Longrightarrow &(n+1, (\xx_1\times\ldots\times \xx_{n+1}, K_1\times\ldots\times K_{n+1}))\in
\kappa^\Delta\circ H( 1^{n+1}0  \langle p_1,\ldots,p_n,p_{n+1}\rangle)\,.
\end{eqnarray*}

Therefore, $H$ realizes $\Pi^{c+}$.

\medskip

(\ref{t6c})
We use the terminology from Definition~\ref{d2}.
 We want to show that the function \\
$((\xx_1,K_1),(\xx_2,K_2),\ldots)\mapsto
(\yy,(K_1\times K_2\times \ldots))$, $Y=\prod_{i=1}^\infty X_i$, is
$([\kappa^\Delta,\kappa^\Delta,\ldots],\kappa^\Delta)$-computable.

By Theorem~\ref{t2}, from $(\xx_1,\xx_2,\ldots)$ we can compute $\yy=\xx_1\times \xx_2\times\ldots$.
It remains to show that from $\kappa^\Delta$-names $\langle t_1,p_1\rangle, \
\langle t_2,p_2\rangle, \ldots)$ of $(\xx_1,K_1),(\xx_2,K_2),\ldots$ we can compute a $\kappa_\yy$-name of the set  $K_1\times K_2\times\ldots\In Y$ (which is compact by the classical Tychonoff theorem~\cite{Eng89}), that is, a list of all $w\in\dom(\nu^{\rm fs}_\yy)$ such that $K_1\times K_2\times \ldots\In \bigcup\nu^{\rm fs}_\yy(w)$.
It suffices to find a Type-2 machine which halts on input $(\langle \langle t_1,p_1\rangle,\langle t_2,p_2\rangle,\ldots\rangle,w)$ such that
$\langle t_i,p_i\rangle\in\dom(\kappa^\Delta)$ and $w\in\dom(\nu^{\rm fs}_\yy)$, iff $K_1\times K_2\times \ldots\In \bigcup \nu_\yy^{\rm fs}(w)$.

Suppose, $\langle t_i,p_i\rangle\in\dom(\kappa^\Delta)$ for $i\in\IN$ and $w\in\dom(\nu^{\rm fs}_\yy)$.
From $w$ we can compute some $n$ and words $v_1,\ldots,v_n$ such that
$\nu_\yy^{\rm fs}(w)=\{\nu_\yy(v_1),\ldots,\nu_\yy(v_n)\}$.
For every $1\leq j \leq n$ we can compute some $m_j$ and words $u_{j1},\ldots,u_{jm_j}$ such that
$ \nu_\yy(v_j)=\nu_1(u_{j1})\times\ldots\times\nu_{m_j}(u_{jm_j})\times X_{m_j+1}\times \ldots$.
Let $m:=\max\{m_j\mid 1\leq j\leq n\}$. Then with Formula~(\ref{f3})
\newpage
\begin{eqnarray*}\nu_\yy(v_j)&=&
 \bigcup\Big\{\nu_1(u_{j1})\times\ldots\times \nu_{m_j}(u_{jm_j})\times \\
&&\hspace{5ex}
\nu_{m_j+1}(u_{j,m_j+1})\times\ldots\times \nu_m(u_{jm})\times X_{m+1}\times\ldots \\
&&\hspace{5ex}\mid u_{j,m_j+1}\in\dom(\nu_{m_j+1}),\ldots,u_{jm}\in\dom(\nu_m)\Big\}\\
&=& \bigcup  \Big\{\nu'_m\langle\ldots\langle u_{j1},u_{j2}\rangle\ldots,
u_{jm_j}\rangle,  u_{j,m_j+1}\rangle \ldots \rangle,u_{jm}\rangle\\
&&\hspace{5ex}\mid u_{j,m_j+1}\in\dom(\nu_{m_j+1}),\ldots,u_{jm}\in\dom(\nu_m)\Big\}
\times X_{m+1}\times\ldots
\end{eqnarray*}

From $w$ a list $p'\in\om$ of all
$\langle\ldots\langle u_{j1},u_{j2}\rangle\ldots,
u_{jm_j}\rangle,  u_{j,m_j+1}\rangle \ldots \rangle,u_{jm}\rangle$
such that $1\leq j\leq n$ and $u_{j,m_j+1}\in\dom(\nu_{m_j+1}),\ldots,u_{jm}\in\dom(\nu_m)$ can be computed.

Then
\[\bigcup \nu^{\rm fs}_\yy(w)= \bigcup\{\nu_m'(w')\mid w'\ll p'\}\times X_{m+1}\times \ldots\,, \]
hence
\[K_1\times K_2\times \ldots\In \bigcup \nu_\yy^{\rm fs}(w)
\iff K_1\times K_2\times \ldots\times K_m\In \bigcup\{\nu_m'(w')\mid w'\ll p'\}\]
By (\ref{t6b}) of this theorem, from $m$ and $\kappa^\Delta$-names of $(\xx_1,K_1),\ldots,(\xx_m,K_m)$ we can compute a $\kappa_{\xx'}$-name $q'$ of
$K_1\times\ldots\times K_m$ (see Definition~\ref{d2} and Formula~(\ref{f3})), which is a list of all $v\in\dom((\nu_m')^{\rm fs})$
such that
\[K_1\times\ldots \times  K_m \In \bigcup (\nu_m')^{\rm fs}(v)\]
Since $K_1\times\ldots\times K_m$ is compact, finitely many $\nu_m'(u)$ with $u\ll p'$ suffice to cover it.
Therefore, $K_1\times K_2\times \ldots\In \bigcup \nu_\yy^{\rm fs}(w)$, iff there are $u_1\ll p', \ldots , u_k\ll p'$ such that
$K_1\times K_2\times \ldots\In \nu_m'(u_1)\cup\ldots\cup \nu_m'(u_k)$, iff there are $u_1\ll p', \ldots , u_k\ll p'$ such that $v:=\iota(u_1)\ldots\iota(u_k)\ll q'$.
There is a Type 2 machine $M$ that halts on input $(p',q')$, iff there are words $u_1\ll p', \ldots , u_k\ll p'$ such that the word $v:=\iota(u_1)\ldots\iota(u_k)\ll q'$.

Let $N$ be a machine which from a $[\kappa^\Delta]^\IN$-name $\langle \langle t_1,p_1\rangle,\langle t_2, p_2\rangle,\ldots\rangle$ of \\ $((\xx_1,K_1),(\xx_2,K_2),\ldots)$ and $w\in\dom(\nu_\yy^{\rm fs})$ first computes $m$ and $p'$, then $q'$ and then  applies $f_M$ to $(p',q')$. This computation halts iff  $K_1\times K_2\times \ldots\In \bigcup \nu_\yy^{\rm fs}(w)$.
\qed

Theorems~\ref{t5} and \ref{t6} can be combined as follows.

\begin{cor}\label{c1}
Theorem~\ref{t6} holds accordingly if $\kappa^\Delta$ is replaced by $\tk^\Delta$.
\end{cor}

\proof By Theorem~\ref{t5} the Cartesian products of sets are computable w.r.t. $\tp^\Delta$. By Theorem~\ref{t6} the Cartesian products of compact sets are computable w.r.t. $\kappa^\Delta$. Therefore, they are computable w.r.t.
$\tp^\Delta\wedge\kappa^\Delta$.  By Lemma~\ref{l1}.\ref{l1c},       $\tp^\Delta\wedge\kappa^\Delta\equiv\tk^\Delta  $
\qed

If $\xx = (X,\tau,\beta,\nu)=\Delta\langle r,s\rangle$, then $r$ supplies information about $\dom(\nu)$ and $s$ supplies information about intersection on the base $\beta$ of the effective topological space.

We observe that the functions $\Pi^{p2}$, $\Pi^{p+} $ and $\Pi^{p\infty}$ from Theorem~\ref{t3} on points and their inverses are computable w.r.t. the multi-representation $\delta$.
If $(\xx,x)\in\delta(p)$ then the information about $\dom(\nu)$ contained in $p\in\dom(\delta)$ is already sufficient to perform the computations.
No additional information from $\Delta$-names about intersection of base elements is needed. The corresponding remark holds for the functions
$\Pi^{s2}$, $\Pi^{s+} $ and $\Pi^{s\infty}$ from Theorem~\ref{t5} on sets.
However, for computing the products  $\Pi^{c2}$, $\Pi^{c+} $ and $\Pi^{c\infty}$ of compact sets the intersection information of the spaces is used.

In Corollaries~\ref{c3} -\ref{c5} let $\xx_1,\xx_2$ be fixed computable topological spaces and let $(\xx_1,\xx_2,\ldots)$ be a
$(\nu_\IN,\Delta)$-computable sequence of (computable) topological spaces.
As a special case of Theorem~\ref{t5} for fixed computable spaces we obtain:

\begin{cor}\label{c3} For subsets $B_i\In X_i$ the following holds.
\begin {enumerate}
\item\label{c3a} The function
$(B_1,B_2)\mapsto B_1\times B_2$  is $([\tp_{\xx_1},\tp_{\xx_2}],\tp_{\xx_1\times\xx_2}))$\bb computable.

\item\label{c3b} For every $n\geq 1$ let $\Pi_n (B_1,\ldots,B_n)= B_1\times\ldots\times B_n$. \\
There is a computable function $f$ such that for every $n\geq 1$,
$q\mapsto f(0^n,q)$ realizes $\Pi_n$ w.r.t
$([\tp_{\xx_1},\ldots,\tp_{\xx_n} ],\tp_{\xx_1\times\ldots\times \xx_n})$.

\item \label{c3c}
The function
$(B_1,B_2,\ldots) \mapsto(B_1\times B_2\times\ldots)$ \\
-- is $([\tp_{\xx_1},\tp_{\xx_2},\ldots ],\tp_{\xx_1\times \xx_2}\times\ldots)$-computable.
\end{enumerate}
\end{cor}

The next corollary is the special case of Theorems~\ref{t7} and~\ref{t3} for fixed computable topological spaces.

\begin{cor}\label{c4} For points $x_i\In X_i$ the following holds.
\begin {enumerate}
\item\label{c4a} $[\delta_{\xx_1},\delta_{\xx_2}]\equiv \delta_{\xx_1\times\xx_2}$.
\item\label{c4b} For every $n\geq 1$ let ${\rm id}_n: (x_1,\ldots,x_n)\mapsto (x_1,\ldots,x_n)$ .
Then there are computable functions $f,g$ such that for every $n\geq 1$,\\
--  $q\mapsto f(0^n,q)$ realizes ${\rm id}_n$ w.r.t
$([\delta_{\xx_1},\ldots,\delta_{\xx_n} ],\delta_{\xx_1\times\ldots\times \xx_n})$ and \\
-- $q\mapsto g(0^n,q)$ realizes ${\rm id}_n$ w.r.t
$(\delta_{\xx_1\times\ldots\times \xx_n},[\delta_{\xx_1},\ldots,\delta_{\xx_n} ])$.
\item \label{c4c}
$[\delta_{\xx_1},\delta_{\xx_2},\ldots] \equiv \delta_{\xx_1\times \xx_2\times\ldots}$
\end{enumerate}
\end{cor}

For the weak multi-representations $\kappa_\xx$ of compact sets we obtain from Theorem~\ref{t6} and Corollary~\ref{c1}.

\begin{cor}\label{c5} For compact subsets $K_i\In X_i$ the following holds.
\begin {enumerate}
\item\label{c5a} The function
$(K_1,K_2)\mapsto K_1\times K_2$  is $([\kappa_{\xx_1},\kappa_{\xx_2}],\kappa_{\xx_1\times\xx_2}))$\bb computable.

\item\label{c35b} For every $n\geq 1$ let $\Pi_n (K_1,\ldots,K_n)= (K_1\times\ldots\times K_n)$ .
There is a computable function $f$ such that for every $n\geq 1$,\\
$q\mapsto f(0^n,q)$ realizes $\Pi_n$ w.r.t
$([\kappa_{\xx_1},\ldots,\kappa_{\xx_n} ],\kappa_{\xx_1\times\ldots\times \xx_n})$.
\item \label{c5c}
The function
$(K_1,K_2,\ldots) \mapsto(K_1\times K_2\times\ldots)$ \\
is $([\kappa_{\xx_1},\kappa_{\xx_2},\ldots ],\kappa_{\xx_1\times \xx_2}\times\ldots)$-computable.
\end{enumerate}
All of this holds accordingly for the strong multi-representations $\tk_\xx$
by minimal covers.
\end{cor}

A last sequence of even less uniform results is obtained from the fact that computable functions map computable points to computable points. For example by Corollary~\ref{c5}, if $K_1$ is $\kappa_{\xx_1}$-computable and $K_2$ is $\kappa_{\xx_2}$-computable then $K_1\times K_2$ is $\kappa_{\xx_1\times\xx_2}$-computable.
We do not list all the other obvious consequences of this kind.

\section{Thanks}
We thank the unknown referees for reading our submission carefully and giving many useful comments.

\bibliographystyle{plain}


\end{document}